\documentclass[
aps,prc,twocolumn,
showpacs,preprintnumbers,
amsmath,amssymb,
floatfix,
superscriptaddress,
]{revtex4-1}

\usepackage{graphicx}         
\usepackage{dcolumn}          
\usepackage{bm}               
\usepackage{color}            
\usepackage{textcomp,gensymb} 
\usepackage[normalem]{ulem}   
\usepackage{xargs}            
\usepackage[colorinlistoftodos]{todonotes}  
\newcommandx{\unsure}[2][1=]{\todo[linecolor=red,backgroundcolor=red!25,bordercolor=red,#1]{#2}}
\newcommandx{\change}[2][1=]{\todo[linecolor=blue,backgroundcolor=blue!25,bordercolor=blue,#1]{#2}}
\newcommandx{\info}[2][1=]{\todo[inline,linecolor=OliveGreen,backgroundcolor=OliveGreen!25,bordercolor=OliveGreen,#1]{#2}}
\newcommandx{\improvement}[2][1=]{\todo[linecolor=Plum,backgroundcolor=Plum!25,bordercolor=Plum,#1]{#2}}

\newcommand{\az}{\ensuremath{^{13}\mathrm{N}}}
\newcommand{\ox}{\ensuremath{^{17}\mathrm{O}}}
\newcommand{\fl}{\ensuremath{^{17}\mathrm{F}}}
\newcommand{\nap}{$^{13}$N($\alpha$,p)$^{16}$O}
\newcommand{\clt}{$^{13}$C($^{7}$Li,t)$^{17}$O}

\newcommand{\zaa}{Astron. Astrophys. }

\newcommand{\zapj}{Astrophys.~J. }
\newcommand{\zapjl}{Astrophys.~J.~Lett. }
\newcommand{\zapjs}{Astrophys.~J.~Supp. }
\newcommand{\zmnras}{MNRAS }

\newcommand{\zcp}{Chinese Physics }
\newcommand{\znp}{Nucl.~Phys. }

\newcommand{\zpr}{Phys.~Rev. }

\newcommand{\znim}{Nucl.~Inst.~and~Meth. }

\newcommand{\zADNDT}{Atomic Data and Nuclear Data Tables }
\newcommand{\znat}{Nature }
\newcommand{\zgca}{Geochimica et Cosmochimica Acta }
\newcommand{\zpasa}{Publications of the Astronomical Society of Australia }

\begin{document}

\title{Evaluation of the \nap\ thermonuclear reaction rate and its impact on the isotopic composition of supernova grains}

\author{A.~Meyer}
   \affiliation{Institut de Physique Nucl\'eaire d'Orsay, UMR8608, IN2P3-CNRS, 
                Universit\'e Paris Sud 11, 91406 Orsay, France}
\author{N.~de~S\'er\'eville}
   \email{deserevi@ipno.in2p3.fr}
   \affiliation{Institut de Physique Nucl\'eaire d'Orsay, UMR8608, IN2P3-CNRS,
                Universit\'e Paris Sud 11, 91406 Orsay, France}
\author{A.~M.~Laird}
   \affiliation{Department of Physics, University of York, York YO10 5DD, 
                United Kingdom}
    \affiliation{NuGrid Collaboration, \url{http://nugridstars.org}}
\author{F.~Hammache}
   \affiliation{Institut de Physique Nucl\'eaire d'Orsay, UMR8608, IN2P3-CNRS, 
                Universit\'e Paris Sud 11, 91406 Orsay, France}
\author{R.~Longland}
   \affiliation{Department of Physics North Carolina State University,
                Raleigh, North Carolina 27695-8202, USA}
   \affiliation{Triangle Universities Nuclear Laboratory, Durham, North
                Carolina, 27708-0308, USA}
\author{T.~Lawson}
   \affiliation{E.~A.~Milne Centre for Astrophysics, Department of Physics and
                Mathematics, University of Hull, HU6 7RX, United Kingdom}
   \affiliation{Konkoly Observatory, Research Centre for Astronomy and Earth Sciences, Hungarian Academy of Sciences, Konkoly Thege Miklos ut 15-17, H-1121 Budapest, Hungary}
   \affiliation{NuGrid Collaboration, \url{http://nugridstars.org}}
\author{M.~Pignatari}
   \affiliation{E.~A.~Milne Centre for Astrophysics, Department of Physics and
                Mathematics, University of Hull, HU6 7RX, United Kingdom}
   \affiliation{Konkoly Observatory, Research Centre for Astronomy and Earth Sciences, Hungarian Academy of Sciences, Konkoly Thege Miklos ut 15-17, H-1121 Budapest, Hungary}
   \affiliation{NuGrid Collaboration, \url{http://nugridstars.org}}
   \affiliation{Joint Institute for Nuclear Astrophysics, Center for the Evolution of the Elements, Michigan State University 640 South Shaw Lane, East Lansing, MI 48824, USA}
\author{L.~Audouin}
   \affiliation{Institut de Physique Nucl\'eaire d'Orsay, UMR8608, IN2P3-CNRS, 
                Universit\'e Paris Sud 11, 91406 Orsay, France}
\author{D.~Beaumel}
   \affiliation{Institut de Physique Nucl\'eaire d'Orsay, UMR8608, IN2P3-CNRS, 
                Universit\'e Paris Sud 11, 91406 Orsay, France}
\author{S.~Fortier}
   \affiliation{Institut de Physique Nucl\'eaire d'Orsay, UMR8608, IN2P3-CNRS, 
                Universit\'e Paris Sud 11, 91406 Orsay, France}
\author{J.~Kiener}
   \affiliation{Centre de Sciences Nucl\'eaires et de Sciences de la Mati\`ere
      (CSNSM), CNRS/IN2P3 et Universit\'e Paris Sud 11, UMR 8609, B\^atiment 
      104, 91405 Orsay Campus, France}
\author{A.~Lefebvre-Schuhl}
   \affiliation{Centre de Sciences Nucl\'eaires et de Sciences de la Mati\`ere
      (CSNSM), CNRS/IN2P3 et Universit\'e Paris Sud 11, UMR 8609, B\^atiment 
      104, 91405 Orsay Campus, France}
\author{M.G.~Pellegriti}
   \affiliation{Institut de Physique Nucl\'eaire d'Orsay, UMR8608, IN2P3-CNRS, 
                Universit\'e Paris Sud 11, 91406 Orsay, France}
   \affiliation{INFN-Sezione di Catania, via Santa Sofia 64
                Catania, Italy}
\author{M.~Stanoiu}
   \affiliation{GSI, Postfach 110552, D-64220 Darmstadt, Germany}
   \affiliation{Horia Hulubei National Institute for Physics and Nuclear 
                Engineering, P.O. Box MG-6, 077125 Bucharest-M\u{a}gurele, Romania}
\author{V.~Tatischeff}
   \affiliation{Centre de Sciences Nucl\'eaires et de Sciences de la Mati\`ere
      (CSNSM), CNRS/IN2P3 et Universit\'e Paris Sud 11, UMR 8609, B\^atiment 
      104, 91405 Orsay Campus, France}

\date{\today}

\begin{abstract}
   \begin{description}
      \item[Background] It has been recently suggested that hydrogen ingestion into the helium shell of massive stars could lead to high $^{13}$C and $^{15}$N excesses when the shock of a core-collapse supernova (CCSN) passes through its helium shell. This prediction questions the origin of extremely high $^{13}$C and $^{15}$N abundances observed in rare presolar SiC grains which is usually attributed to classical novae. In this context the \nap\ reaction plays an important role since it is in competition with $^{13}$N $\beta^+$-decay to $^{13}$C.
      \item[Purpose] The \nap\ reaction rate used in stellar evolution calculations comes from the Caughlan \& Fowler compilation with very scarce information on the origin of this rate and with no associated uncertainty. The goal of this work is to provide a recommended \nap\ reaction rate, based on available experimental data, with a meaningful statistical uncertainty.
      \item[Method] Unbound nuclear states in the \fl\ compound nucleus were studied using the spectroscopic information of the analog states in \ox\ nucleus that were measured at the Tandem-Alto facility using the \clt\ alpha-transfer reaction. Alpha spectroscopic factors were derived using a Finite-Range Distorted-Wave Born Approximation (FR-DWBA) analysis. This spectroscopic information was used to calculate a recommended \nap\ reaction rate with meaningful uncertainty using a Monte Carlo approach.
      \item[Results] The \nap\ reaction rate from the present work is found to be within a factor of two of the previous evaluation in the temperature range of interest, with a typical uncertainty of a factor $\approx2-3$. The source of this uncertainty has been identified to come from the three main contributing resonances at $E_r^{c.m.} = 221$, 741 and 959~keV. This new error estimation translates to an overall uncertainty in the $^{13}$C production of a factor of 50 when using the lower and upper reaction rates in the conditions relevant for the \nap\ activation.
      \item[Conclusions] The main source of uncertainty on the re-evaluated \nap\ reaction rate currently comes from the uncertain alpha-width of relevant \fl\ states.
   \end{description}
\end{abstract}

\maketitle

\section{Introduction}
Abundance measurements 
of isotopes and elements in stars provide a fundamental diagnostic for stellar evolution and internal stellar conditions.
Theoretical predictions from stellar models can be directly compared with
observations of 
very old stars
\cite{yoon:18}, or with 
evolved stars of any age including the Sun by using
galactical chemical evolution simulations
\cite{gibson:03,kobayashi:11,mishenina:17}. Specific information about
individual stars and 
supernova explosions
can be obtained, by e.g., observing abundance signatures from supernova remnants
\cite{grefenstette:14,yamaguchi:15},
or by measuring abundances in single presolar grains found in meteorites.
Presolar grains condensed around old dying stars
like supernovae and Asymptotic Giant Branch stars just before the formation of
the Sun, and 
then were trapped in meteorites
formed in the early solar system. Pristine isotopic abundances in single
presolar grains, therefore,
carry the signature of their parent stars \cite{zinner:14}.
Isotopic ratios that are measured in single presolar
grains can be used as a constraint to map stellar structure properties.
Carbon-rich presolar grains from core-collapse supernovae
provide fundamental insights about the supernova explosion, and about the
progenitor massive star, specifically from the
He-burning layers \cite{pignatari:13,zinner:14}.
Data coming from presolar dust like SiC grains of Type X
\cite{besmehn:03}, Type C \cite{pignatari:13b}
and low-density graphites \cite{amari:90} challenge theoretical
supernova models, highlighting their limitations and providing new puzzles to
solve. Nuclear reaction rates relevant in these conditions
are crucial ingredients of these models to define final stellar abundances.

Among the different types of presolar SiC grains, putative nova grains represented,
for many years, an unsolved challenge for stellar models
\cite{nittler:05}. Nova grains show high excesses of isotopes $^{13}$C
and $^{15}$N compared to the solar composition, that can
be explained by the hot CNO cycle during typical nova conditions
\cite{jose:07,denissenkov:14}.
However, some of the nova grains also showed $^{44}$Ca excess, which can only be
explained as radiogenic contribution of the radioactive
isotope $^{44}$Ti. $^{44}$Ti can be made in supernovae but not in novae, while
standard supernova models were not able to explain
the observed $^{13}$C and $^{15}$N abundances \cite{nittler:05}.
A realistic solution for this conundrum was
provided by \cite{pignatari:15}, using new supernova
models where fresh hydrogen was ingested in the He-rich stellar layers
of massive star progenitors, just before the supernova explosion. The
nucleosynthesis obtained in the H-ingestion event, and the mixture of
explosive He-burning and H-burning yields generated by the following SN shock in
the He-rich layers, provide the conditions to generate sufficient
$^{13}$C and $^{15}$N abundances to explain measurements in putative nova
grains. Typical temperatures ranging between 0.4 and 1 GK in the SN shock are achieved depending on the amount of H available in He-rich layers. However
multi-dimensional hydrodynamics models are required to quantitatively study the stellar structure response and nucleosynthesis following H-ingestion events. While models of this kind exist for ingestion of H into the He shell in AGB stars, post-AGB stars and in Rapidly Accreting WDs \citep[e.g.,][]{stancliffe:11,herwig:14,denissenkov:19}, the first hydrodynamics simulations are only recently becoming available for massive stars \citep[][]{clarkson:18}. For this reason, the nucleosynthesis analysis of Ref.~\cite{pignatari:15} took into account different SN explosion energies and a large range of H concentration left after the ingestion. 
While a new generation of stellar models for massive stars informed from multi-dimensional hydrodynamics simulations are needed to drive more definitive conclusions, \cite{pignatari:15} showed that the production of $^{13}$C and $^{15}$N in He-rich layers consistent with the abundance pattern in putative nova grains is obtained for a wide combination of SN explosion energies and H concentration. 
In these models, during the SN explosion, the reaction
$^{13}$N($\alpha$,p)$^{16}$O is efficiently activated.
$^{13}$N is made by proton capture on $^{12}$C. The accumulation of $^{13}$N in
the He shell will determine how much radiogenic $^{13}$C will be ejected by
the explosion. On the other hand, $^{13}$N($\alpha$,p)$^{16}$O is depleting part
of the $^{13}$N made, producing instead $^{16}$O.

\begin{figure}[!htpb]
  \includegraphics[width=0.5\textwidth]{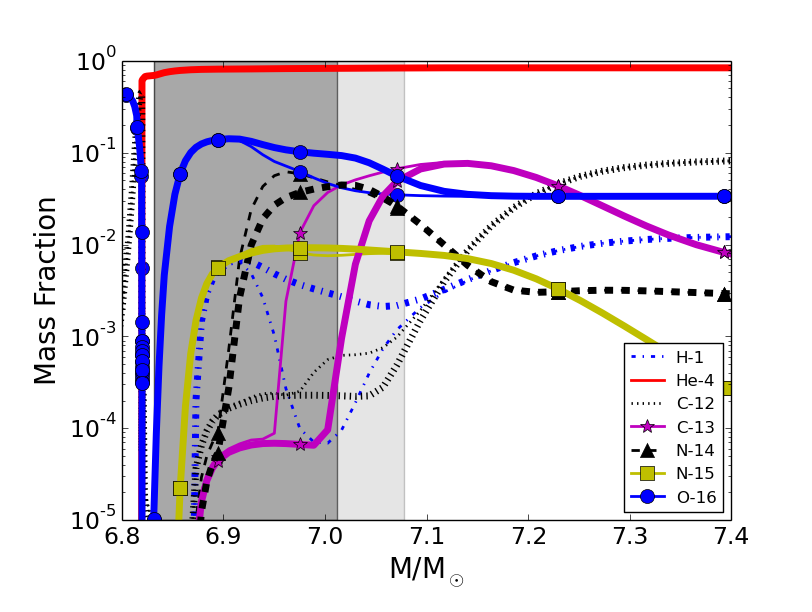}
  \caption{\label{f:impact}
     (Color online) Isotopic abundances in the He-shell ejecta of a 25~$M_\odot$
     supernova model. Thick (thin) lines correspond to a \nap\ reaction rate variation 
     by a factor of five up (down), respectively. Shaded area identify the O-rich zones (or O-Nova zone, with C/O$<$1) in the He shell region, where the heavy-grey (light-grey) area is obtained by using the higher (lower) \nap\ reaction rate.}
\end{figure}

The \nap\ thermonuclear reaction rate used in stellar models~\cite{pignatari:15} comes from 
the Caughlan \& Fowler~\cite{CF88} (hereafter CF88) compilation. The impact of a 
variation of the \nap\ reaction rate by an arbitrary factor of five with respect to the 
CF88 rate has been investigated and the results for decayed abundances are shown in Fig.~\ref{f:impact}, using the stellar simulations by \cite{pignatari:15}.
The largest abundance variation is shown for H, $^{13}$C and $^{16}$O, when the temperature peak of the SN shock is between 0.5 GK and 0.7 GK.  A higher $^{13}$N($\alpha$,p)$^{16}$O reaction rate destroys $^{13}$N, producing more $^{16}$O. Therefore, the abundance of $^{13}$C, from the $^{13}$N decay, decreases. The higher abundance of H is also due to a stronger activation of the ($\alpha$,p) channel. Since the H reservoir is affected, the $^{13}$N($\alpha$,p)$^{16}$O rate might potentially affect the efficiency of other proton capture reactions. In the final part of this work we will discuss this in more detail.

The thermonuclear \nap\ reaction rate given in the CF88 compilation comes from the reverse $^{16}$O(p,$\alpha$)$^{13}$N reaction. However it is not clear from the CF88 compilation (and references therein) what is the origin of the nuclear data used to derive the $^{16}$O(p,$\alpha$)$^{13}$N reaction rate; moreover no reaction rate uncertainty is given.
A compilation of $^{16}$O(p,$\alpha$)$^{13}$N excitation functions can be found 
in Ref.~\cite{Tak03} and some of the reported works~\cite{Ner73,Gru77} give reaction 
rates for temperatures $T_9$~$>$~1.4 ($T_9~\equiv~T(K)/10^9$).
Unfortunately this is higher than the temperature range of interest $T_9 = 0.4 - 1$ when the SN
shock crosses the He-shell.
The first estimate of the thermonuclear \nap\ reaction rate was given by Wagoner et al.~\cite{Wag67,Wag69} based on the formalism for non-resonant reactions~\cite{Bah66}, but no details are given on the origin of the numerical values used in the analytical formula of the reaction rate.
Another estimate of the thermonuclear \nap\ reaction rate based on the Hauser-Feshbach 
model can be found in the STARLIB library~\cite{Sal13}. However the use of such a 
nuclear model for a low mass number ($A = 17$) nuclide with low level density is 
questionable and an uncertainty of a factor of 10 has been associated to the \nap\ 
reaction rate~\cite{Sal13}.
Given this situation a re-evaluation of the thermonuclear \nap\ reaction rate
including a meaningful statistical uncertainty is necessary to constrain the effect of this rate on the final $^{13}$C abundance.

The evaluation of the \nap\ reaction rate in the temperature range of interest $T_9 = 
0.4 - 1$ requires a detailed knowledge of the structure of the compound nucleus \fl\ 
within around 2.5~MeV above the $^{13}$N+$\alpha$ threshold. State energies are known 
though with a relatively large uncertainty of a few tens of keV~\cite{Til93}. Spins and 
parities are known in most cases and the total widths are known experimentally~\cite{Til93}. 
Given that the $^{13}$N+$\alpha$ threshold ($S_{\alpha+\az}$~=~5818.7~(4)~keV) is much higher than the
$^{16}$O+p threshold ($S_p$~=~600.27~(25)~keV), the states in the region of interest decay mainly by proton emission, so that $\Gamma_p \approx \Gamma_{tot}$.
Their contribution to the reaction rate is 
therefore directly proportional to their unknown alpha-particle widths. This paper
provides an evaluation of the alpha-particle widths of \fl\ states based on the
properties of \ox\ analog states when a pairing connection exists.

The goal of this work is to determine statistically meaningful thermonuclear
rates for the \nap\ reaction. Unfortunately a direct measurement of this reaction 
cross section is not currently feasible with existing $^{13}$N beam intensity,
and therefore we rely on an indirect approach. We first report on the analysis of 
\clt\ alpha transfer reaction measurement in order to determine the alpha spectroscopic factors of \ox\ analog states of \fl\ (Sec.~\ref{experiment}).
Under the mirror symmetry assumption, spectroscopic information for the analog
\fl\ states is then derived (Sec.~\ref{fluor17}) and further used to evaluate
thermonuclear rates and rate uncertainties (Sec.~\ref{rates}). Finally, the impact
of the new \nap\ reaction rate in the hydrogen ingestion scenario in massive
stars is explored (Sec.~\ref{astro}).

\section{STUDY OF THE \clt\ TRANSFER REACTION} \label{experiment}
\subsection{Experimental procedure}
The \clt\ reaction measurement~\cite{Pel08} was performed at the Tandem-ALTO facility in Orsay, France. Experimental details can be found in~\cite{Pel08} and the most relevant 
information for the present study is recalled here. A $^{7}$Li$^{3+}$ beam of
about 100~enA was accelerated by the 15~MV Tandem to an energy of 34~MeV. The
beam impinged on a self-supporting enriched (90\%) $^{13}$C target of 
80(4)~$\mu$g/cm$^2$ located at the object focal plane of an Enge Split-Pole
magnetic spectrometer~\cite{Spe67}. Light reaction products were momentum
analyzed and focused on the focal-plane detection system~\cite{Mar75}, and 
tritons were readily distinguished
from deuterons using the energy loss and 
magnetic rigidity measurements. The tritons were detected at
eleven angles between 0\degree\ and 33\degree\ in the laboratory frame.
The unreacted beam was detected inside the reaction chamber by a Faraday cup 
at 0\degree\ recording the accumulated charge of each run.

\subsection{Data reduction}
\begin{figure}[!htpb]
  \includegraphics[width=0.52\textwidth]{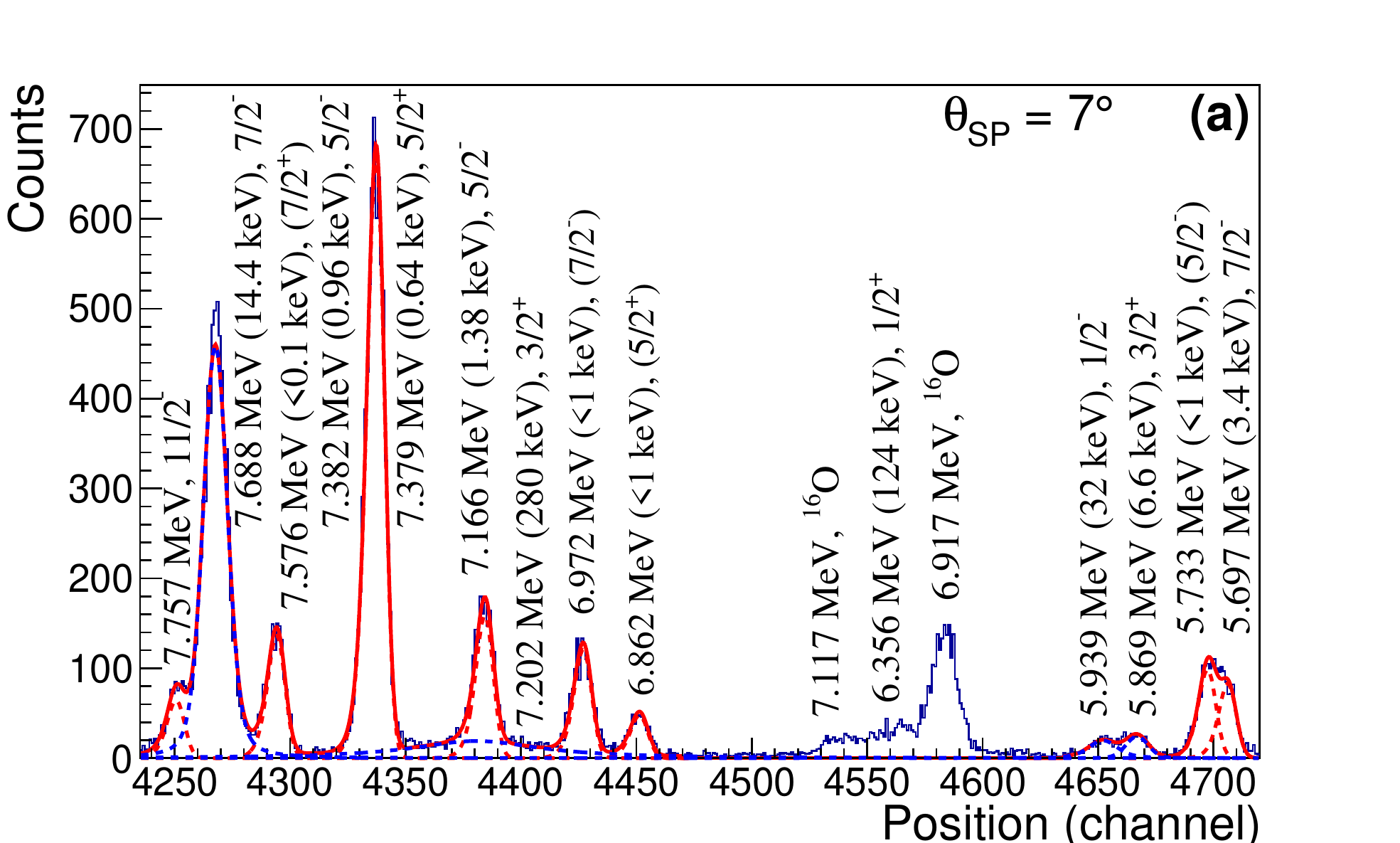} \\
  \includegraphics[width=0.52\textwidth]{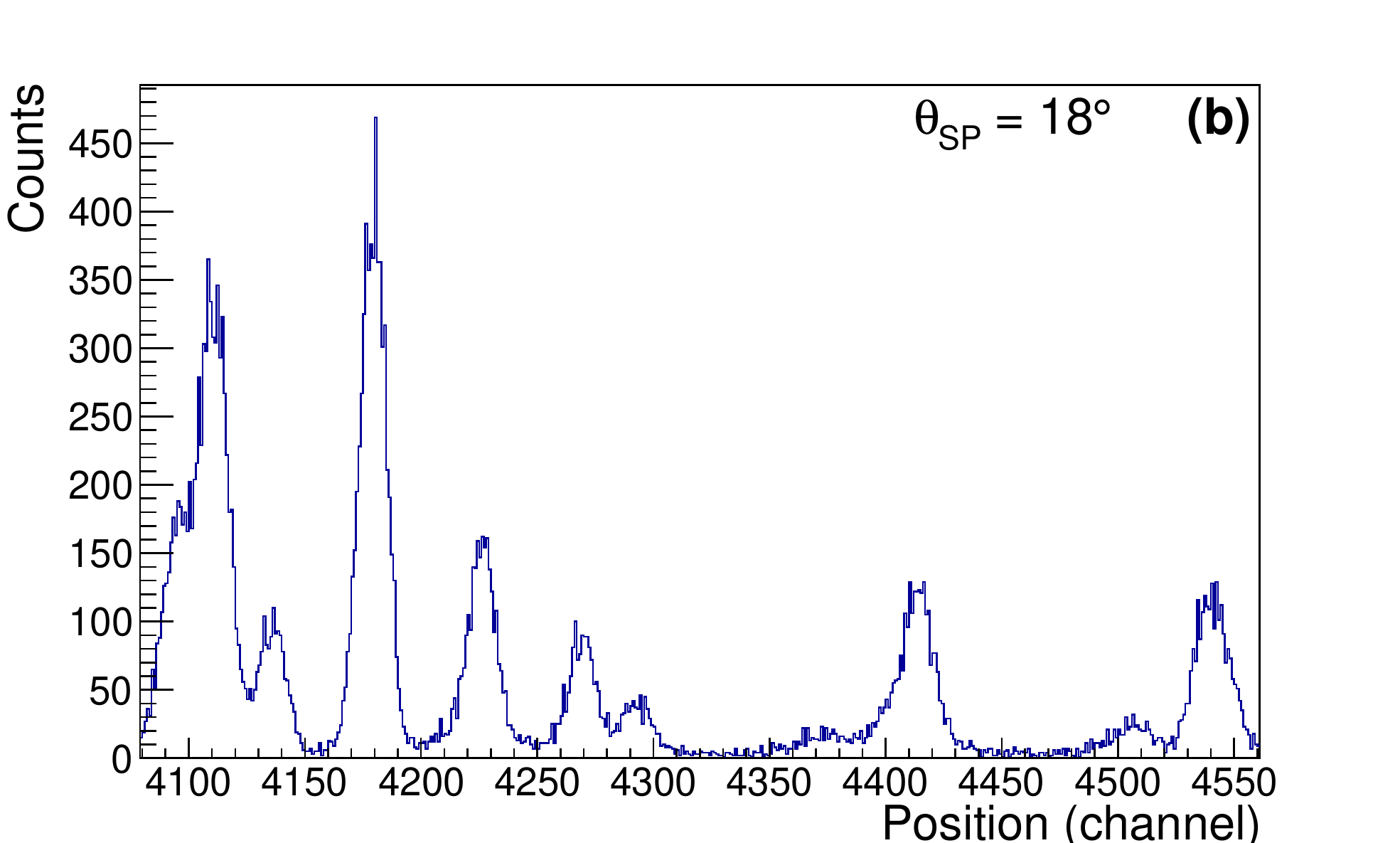}
  \caption{\label{f:tritons}
     (Color online) Triton magnetic rigidity spectrum at spectrometer
     angle of 7\degree\ and 18\degree corresponding to an incident charge of 585~$\mu$C and 1155~$\mu$C, respectively. 
     Excitation energies in \ox\ between 5.6 and 7.8~MeV are covered. All 
     triton peaks correspond to known \ox\ states unless this is indicated. 
     For the data at 7\degree, the best fit of the spectrum is shown (solid line), together with  
     individual contributions (dashed lines) for narrow states (red) and states with known widths (blue). Labeled energies, total widths and spin-parities are from the last NNDC compilation~\cite{Til93}, except for the broad state at 7.202~keV whose width is from the present analysis.
     }
\end{figure}

After selection, triton spectra of the focal-plane position were obtained for
each spectrometer angle and the case of 7\degree\ and 18\degree\ are shown in 
Fig.~\ref{f:tritons}. Apart from the two triton contamination peaks associated 
to $^{16}$O states at 6.917 and 7.117~MeV, all peaks could be identified with 
known \ox\ states. This identification relies on two considerations: the use of the focal-plane detector calibration and the kinematics of the \clt\ reaction.

The calibration of the focal-plane detector was performed using a $^{nat}$C target and narrow well-isolated $^{16}$O states populated from the $^{12}$C($^{7}$Li,t)$^{16}$O reaction. The relation between the radius of curvature and the focal-plane position was obtained and the calibration deduced after fitting this relation by a one-degree polynomial function. The calibration was then applied to the raw data and the magnetic rigidity of the observed triton peaks matched the expectation from the energy of \ox\ states.

Comparison of the triton peaks at the spectrometer angles of 7\degree\ and 18\degree\ shows that the relative position of the peaks is the same. This behaviour confirms that the triton peaks correspond to excited states belonging to the same nucleus. It was checked that the experimental difference of magnetic rigidity between angles for a same state was following the \clt\ kinematics. This again supports the identification of triton peaks as \ox\ excited states. Any two-body reaction occurring on nuclei different from $^{13}$C (e.g. contaminants in the target) will produce triton peaks that will have a different kinematic dependence than \ox\ states. This is the case for the peaks associated to $^{16}$O states at 6.917 and 7.117~MeV states, which are moving toward the \ox\ state at 6.862~MeV as the detection angle is increasing.

The triton magnetic rigidity spectra were independently analyzed using a least-squares fit of multiple Gaussian and Voigt functions at each detection angle, and the best fit was obtained. The Gaussian function was used to describe \ox\ states having natural widths much smaller than the  experimental resolution of $\approx$~50~keV (FWHM, center of mass). A common width was used as a free parameter in the fitting procedure. The Voigt function was used to describe triton peaks associated to \ox\ states at 5.697-, 5.869-, 5.939-, 7.202- and 7.688-MeV, which have a sizeable total width. The natural width was kept as a fixed parameter in the Lorentzian component of the Voigt function while the width of the Gaussian component was the same free parameter as for the Gaussian used to describe the narrow states. 
The natural width of the \ox\ state at 7.202~MeV was determined from the present data (see below). The magnetic rigidity region around the \ox\ state at 6.356~MeV and the two $^{16}$O contamination states were excluded from our fitting procedure since results concerning this energy region have already been reported~\cite{Pel08}. 
The best fit of the triton magnetic rigidity spectrum obtained at a spectrometer angle of 7\degree\ is represented in Fig.~\ref{f:tritons}.
The states at 5.697 and 5.733~MeV were not included in the fitting procedure for the higher detection angle because they were hindered by an $^{16}$O contamination state.

\begin{figure}[!htpb]
  \includegraphics[width=0.52\textwidth]{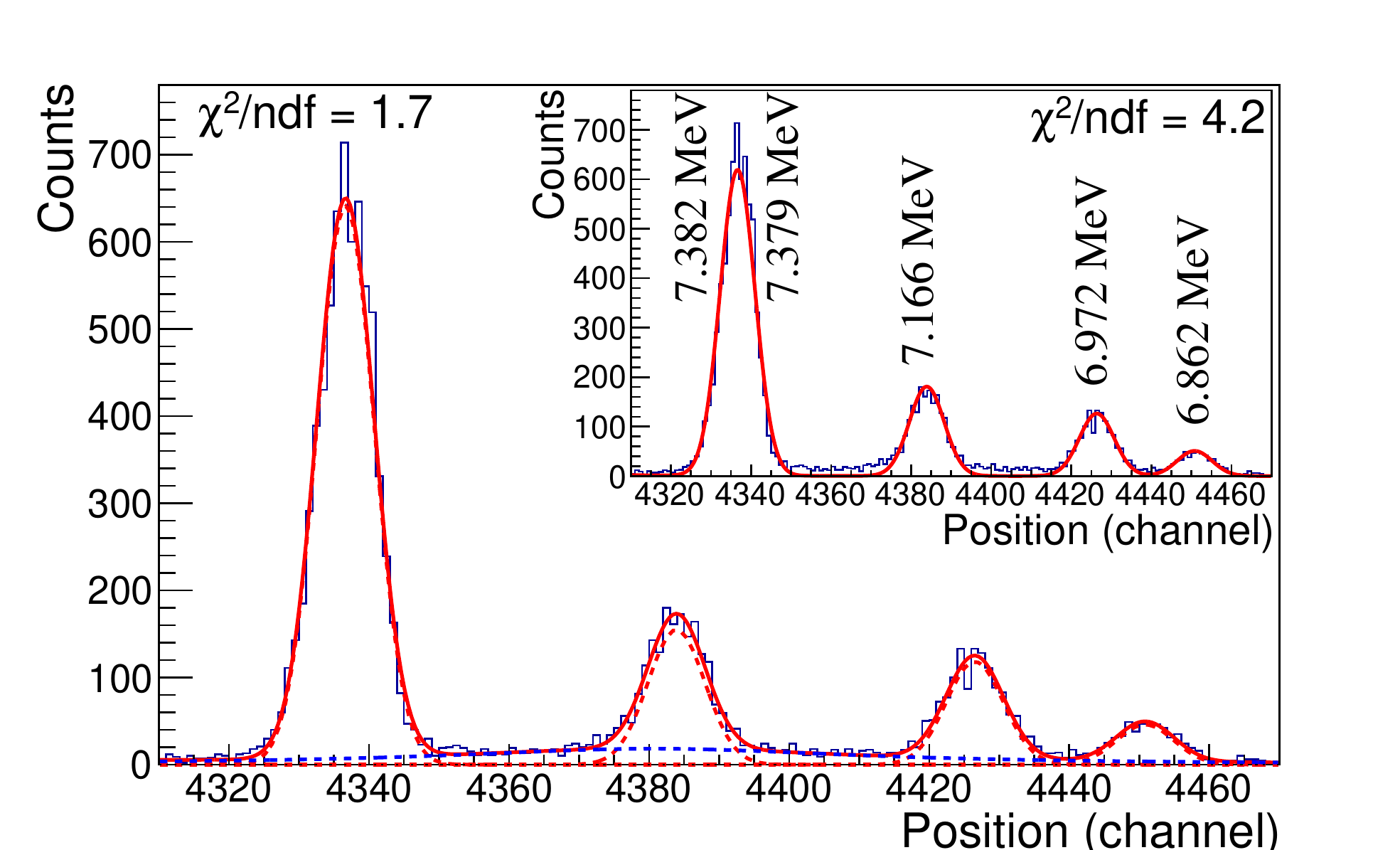}
  \caption{\label{f:broad}
     (Color online) Triton magnetic rigidity spectrum at spectrometer
     angle of 7\degree\ zoomed in the \ox\ excitation energy region between 6.8 and 7.4~MeV. The broad \ox\ state at 7.202~MeV is included in the fitting procedure and its contribution is represented by the dashed blue line. The insert shows a fit of the same data without the inclusion of the broad state. Reduced chi-squares are also given.
     }
\end{figure}

A close-up of the excitation energy region between 6.8 and 7.4~MeV is shown in Fig.~\ref{f:broad} where the contribution of the \ox\ state at 7.202~MeV is represented by dashed blue line. The insert in Fig.~\ref{f:broad} corresponds to the fitting case when the broad state is not taken into account. The reduced chi-square is much better when the broad state is included ($\chi^2$/ndf=1.7) than without broad state ($\chi^2$/ndf=4.2), which strongly supports the observation of the 7.202~MeV state in the present data.
Several values of the total width of the broad \ox\ state at 7.202~MeV can be found in the literature, ranging from 280~(30)~keV~\cite{Joh73, Til93} to 400~(30)~keV~\cite{Lis66}, while a recent measurement reports 262~(7)~keV~\cite{Fae15}. The natural width of the 7.202~MeV state was therefore kept as a free parameter in the fitting procedure described above and a value of 313~(22)~keV was found after averaging over the first eight smaller spectrometer angles. 
Our result agrees within 1-$\sigma$ with the adopted value from Ref.~\cite{Joh73, Til93} and within 2-$\sigma$ with the two other values available in the literature~\cite{Lis66, Fae15}.

\subsection{Angular distributions and DWBA analysis}
\begin{figure*}[!htpb]
  \includegraphics[width=1.02\textwidth]{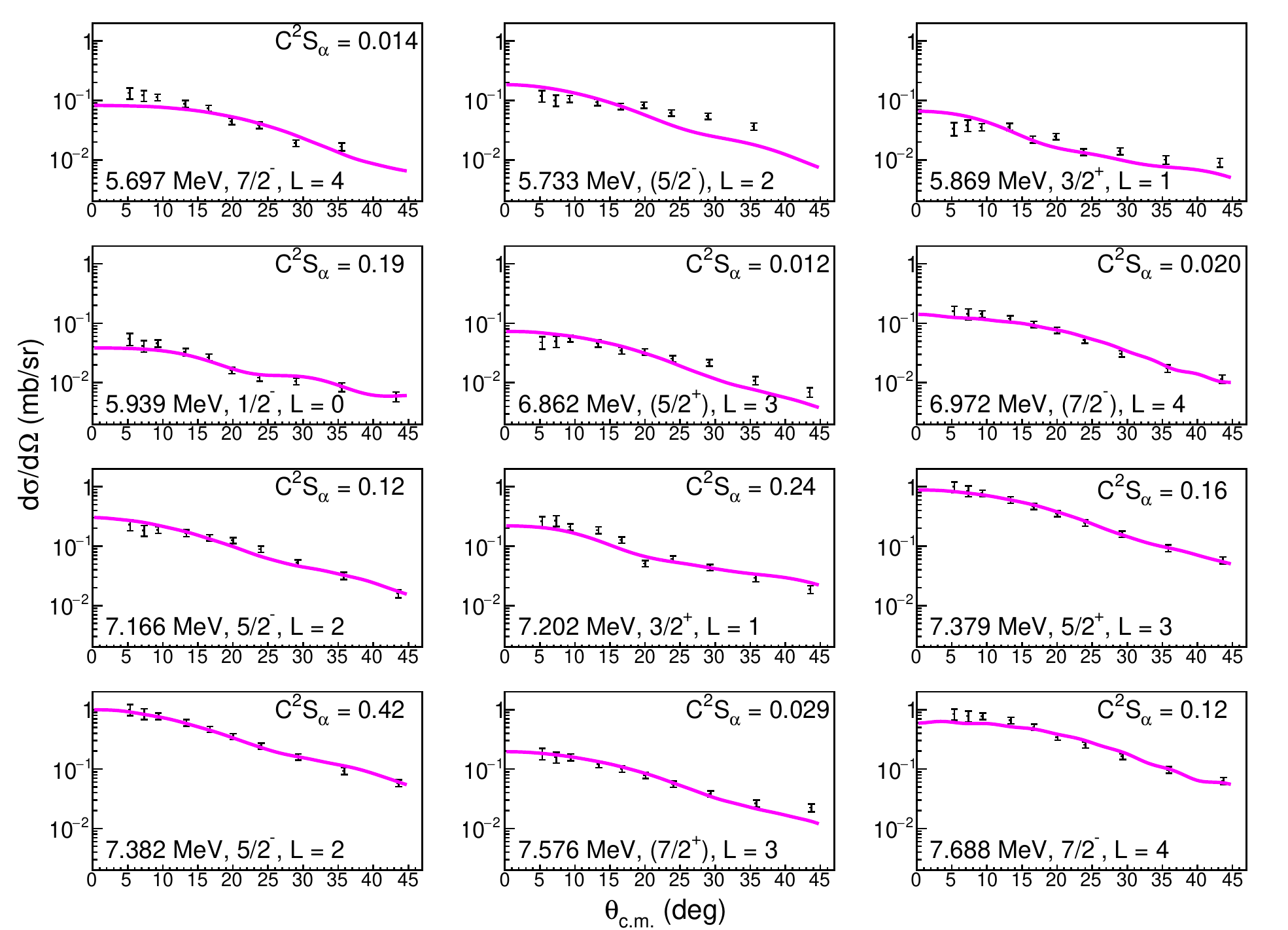}
  \caption{\label{f:dsig}
     (Color online) Experimental differential cross sections of \ox\ states 
     populated with the \clt\ reaction at 34~MeV. Solid lines represent
     finite-range DWBA calculations normalized to the data.}
\end{figure*}

The differential cross sections corresponding to populated \ox\ states were
calculated from the triton yield determined at each spectrometer angle 
$Y_t(\theta_{lab})$ using the following formula
\begin{equation}
  \left(\frac{d\sigma}{d\Omega}\right)_{c.m.}(\theta_{c.m.}) = 
  \frac{Y_t(\theta_{lab})}{Q(\theta_{lab}) N_{target} \Delta\Omega_{lab}} J(\theta_{lab})
  \label{e:dsig}
\end{equation}
where $Q(\theta_{lab})$ is the accumulated charge at each angle, $N_{target}$
is the number of $^{13}$C atoms per unit area, $\Delta\Omega_{lab}$ is the 
Split-Pole solid angle, and $J(\theta_{lab})$ is the Jacobian for the laboratory 
to center-of-mass transformation of the \clt\ reaction at each spectrometer angle. 
The differential cross sections are shown in Fig.~\ref{f:dsig} together with 
Finite-Range Distorted-Wave Born Approximation (FR-DWBA) calculations 
performed with the FRESCO code~\cite{FRESCO}.

We follow the prescription from Ref.~\cite{Pel08} for the choice of optical 
potential  parameters and for the overlap between the $\alpha$+t and $^{7}$Li
systems. Several combinations of entrance and exit optical potential parameters 
have been tested as inputs of the DWBA calculations~\cite{Ser17}. The best compromise for 
describing differential cross sections for all \ox\ states at the same time 
was obtained with the potential III from Ref.~\cite{Sch73} for the
$^{13}$C+$^{7}$Li entrance channel, and with the potential I.a from 
Ref.~\cite{Gar73} for the t+$^{17}$O exit channel. Concerning the $\alpha$-wave 
function in \ox, the depth of a Woods-Saxon potential ($r$ = 4~fm and $a$ = 
0.76~fm) was adjusted to reproduce the known $\alpha$-separation energy for 
each state. The number of radial nodes $N$ (including the origin) of the $\alpha$-wave function in 
\ox\ was set using the usual oscillator energy conservation rule~\cite{Mos59}
when the number of quanta in the relative motion $Q=2(N-1)+L$
is equal to 6 for negative-parity states and 7 for positive-parity states.
This can be linked to 2p-1h and 3p-2h shell model configurations for negative-
and positive-parity states, respectively, as suggested by theoretical 
calculations~\cite{Bro66} for \ox\ states of high excitation energies. 
Note that the shape of the angular distribution calculated by the DWBA
model shows very little sensitivity to the number of nodes $N$. In the case
of negative-parity states we indeed considered calculations with $Q=8$
which could be associated to the possible 4p-3h configuration, and as 
expected the shape of the calculated angular distributions were very similar
though the $Q=6$ case slightly better described the data.

As can be seen in Fig.~\ref{f:dsig} a very good agreement is observed between
normalized FR-DWBA calculations and the data in most cases. This supports a 
single step direct mechanism for the population of \ox\ states using the \clt\
reaction; the only exception being for the two \ox\ states at 5.733 and 
5.869~MeV.
This is not surprising since their experimental 
differential cross sections vary less strongly as a function of the center-of-mass angle, which suggests that these states are significantly populated by the triton evaporation of the compound nucleus $^{20}$F or by a multiple step reaction mechanism.

The normalization factor between the experimental and DWBA differential cross
sections for a given state is equal to the product of the \ox\ alpha spectroscopic 
factor ($C^2S_\alpha$) and the square of the overlap between the $\alpha$+t 
and $^{7}$Li systems ($S_\alpha^{^{7}\mathrm{Li}}$). We used 
$S_\alpha^{^{7}\mathrm{Li}} = 1$ in the present work following the prescription from Ref.~\cite{Pel08}. Determination of alpha spectroscopic factors for unbound \ox\ 
states follows the prescription given in Ref.~\cite{Bec78}. The calculation of 
the $\alpha$-wave function for unbound \ox\ states used form factors obtained 
with the $\alpha$-cluster bound at 0.1~MeV. This should be suitable for states 
associated to large transferred angular momentum ($L\geq2$) since the 
$\alpha$-cluster is quasi-bound due to the large centrifugal barrier. In case 
of lower transferred angular momentum such as for the \ox\ state at 7.202~MeV 
($L=1$) the calculation was performed at several $\alpha$-binding energies 
approaching zero and the DWBA cross section was extrapolated to the actual 
$\alpha$-separation energy (see Fig.~3 in Ref.~\cite{Ser17} for an example).

For unbound states the alpha partial width can be deduced from the 
corresponding spectroscopic factor using the following formula~\cite{Ili08}
\begin{equation}
  \Gamma_\alpha = 2P_L(r,E)\frac{\hbar^2r}{2\mu} C^2S_\alpha\ |\phi(r)|^2, 
  \label{e:width}
\end{equation}
where $\mu$ is the reduced mass for the $\alpha$+$^{13}$C system, $P_L(r,E)$ 
is the penetrability of the Coulomb and centrifugal barriers for transferred
angular momentum $L$, and $|\phi(r)|$ is the radial part of the 
$\alpha$+$^{13}$C wave function. Eq.~\ref{e:width} has been evaluated at the
interaction radius $r=7.5$~fm where the $\alpha$+$^{13}$C wave function reaches
an asymptotic behavior~\cite{Ser17}.

The parameters used in the FR-DWBA analysis and the results from the present work are presented in Tab.~\ref{t:c2s}. Comparison with alpha widths determined from previous experimental work reported in the last NNDC compilation~\cite{Til93} is also provided. A very good agreement is found between our results and the literature, typically within a factor of two. The only noticeable difference is for the \ox\ state at 7382.2~keV which is part of an unresolved doublet with the 7379.2~keV state in the present experiment. If we assume that all the strength is on the 7379.2~keV state, we find an alpha width in very good agreement with NNDC~\cite{Til93}. On the other hand, if we assume that all the strength is on the 7382.2~keV state, our determination of the alpha width is about 50 times larger than the one reported in NNDC~\cite{Til93}. This indicates most probably that the \ox\ state at 7379.2~keV has been preferentially populated in the present experiment. 

Comparison with alpha widths determined from works using $\mathcal{R}$-matrix analysis of the $^{13}$C($\alpha$,n)$^{16}$O reaction~\cite{Hei08, Say00} is also provided in Tab.~\ref{t:c2s}. A good agreement is obtained for excitation energies greater than 7~MeV with the exception of the \ox\ state at 7382.2~keV as explained before. Below 7~MeV there is no $^{13}$C+$\alpha$ experimental data which can be used to constrain the alpha widths of \ox\ states. This explains the difference between our results and those of Ref.~\cite{Hei08} which come from an extrapolation of the cross section measured at higher energies.

\begin{table*}[!htpb]
  \caption{\label{t:c2s}
     Alpha spectroscopic factors and widths for $^{17}$O states obtained 
     from the present analysis. Comparison with alpha widths from the 
     literature is provided.}
  \begin{ruledtabular}
  \begin{tabular}{cccccccccc}
    \multicolumn{3}{c}{NNDC~\cite{Til93}} & \multicolumn{3}{c}{Present work} & \multicolumn{2}{c}{Heil et al.~\cite{Hei08}} & \multicolumn{2}{c}{Sayer et al.~\cite{Say00}} \\ \cline{1-3} \cline{4-6} \cline{7-8} \cline{9-10}
    $E_x$ & $J^\pi$ & $\Gamma_\alpha$ & $N, L$\footnotemark[1] & $C^2S_\alpha$ & $\Gamma_\alpha$\footnotemark[2] & $E_x$ & $\Gamma_\alpha$ & $E_x$ & $\Gamma_\alpha$  \\
    (keV) &         &  (keV)          &        &       & (keV)        & (keV)          & (keV) & (keV) & (keV) \\ \hline
    5697.3 (4)  &  7/2$^-$  &        & 2, 4 & 0.014 & &  5696.7 & 2.4$\times$10$^{-11}$ & 5696.7 & \\
    5732.8 (5)  & (5/2$^-$) &        & 3, 2 &       & &  5733.5 & 4.1$\times$10$^{-9}$  & 5732.3 & \\
    5869.1 (6)  &  3/2$^+$  &        & 4, 1 &       & &  5868.4 & -4.1$\times$10$^{-4}$ & 5868.7 & \\
    5939   (4)  &  1/2$^-$  &        & 4, 0 & 0.19  & &  5923.2 & 5.5$\times$10$^{-9}$  & 5932.0 & \\
    6356   (8)  &  1/2$^+$  &        & 4, 1 & 0.29\footnotemark[3]  & $13.5\pm6.6$\footnotemark[3] &  6379.5 & 1.7$\times$10$^{-54}$ & 6380.2 & \\
    6862   (2)  & (5/2$^+$) &        & 3, 3 & 0.012 & 1.1$\times$10$^{-7}$ &  6829.8 & 1.1$\times$10$^{-6}$  & 6860.7 & \\
    6972   (2)  & (7/2$^-$) &        & 2, 4 & 0.020 & 8.2$\times$10$^{-8}$ &  6936.2 & 3.3$\times$10$^{-6}$  & 6971.9 & \\
    7165.7 (8)  &  5/2$^-$  & 0.0033 & 3, 2 & 0.12  & 3.4$\times$10$^{-3}$ &  7164.6 & 4.3$\times$10$^{-3}$  & 7164.6 & 0.009 \\
    7202  (10)  &  3/2$^+$  & 0.07   & 4, 1 & 0.24  & 7.3$\times$10$^{-2}$ &  7247.7 & 0.14    & 7239.1 & 0.17 \\
    7379.2 (10) &  5/2$^+$  & 0.01   & 3, 3 & 0.16\footnotemark[4] & 8.0$\times$10$^{-3}$ &  7377.9 & 0.011   & 7378.2 & 0.02 \\
    7382.2 (10) &  5/2$^-$  & 0.003  & 3, 2 & 0.42\footnotemark[4] & 0.131 &  7380.7 & 2.9$\times$10$^{-3}$  & 7380.8 & 0.007 \\
    7559   (20) &  3/2$^-$  & 0.08   & &      & &  7475.2 & 0.027   & 7446.9 & 0.026 \\
    7576    (2) & (7/2$^+$) &        & 3, 3 & 0.029 & 7.3$\times$10$^{-3}$ &         &         &        &  \\
    7688.2  (9) &  7/2$^-$  & 0.01   & 2, 4 & 0.12  & 3.3$\times$10$^{-3}$ &  7686.0 & 0.011   & 7686.9 & 0.026 \\
  \end{tabular}                                                                
  \end{ruledtabular}
  \footnotetext[1] {The quantities $N$ and $L$ are the radial nodes (including the origin) and orbital 
  angular momentum assigned to the center of mass motion of the 
  $\alpha$-cluster in $^{17}$O.}
  \footnotetext[2] {$\Gamma_\alpha = 2P_l(a,E)\frac{\hbar^2a}{2\mu} 
  C^2S_\alpha\; |\phi(a)|^2$ with $|\phi(a)|$ being the radial part of the 
  $^{13}$C+$\alpha$ wave function evaluated at the channel radius $a =7.5$~fm (see text).}
  \footnotetext[3] {From~\cite{Pel08}, the reduced width $\gamma^2_\alpha$ is given instead of $\Gamma_\alpha$.}
  \footnotetext[4] {This doublet is not resolved experimentally so the deduced 
  spectroscopic factor assumes all the strength is on one or the other state.}
\end{table*}

\section{\label{fluor17} RESONANCE PARAMETERS IN \fl}
\begin{figure*}[!htpb]
  \includegraphics[width=0.82\textwidth]{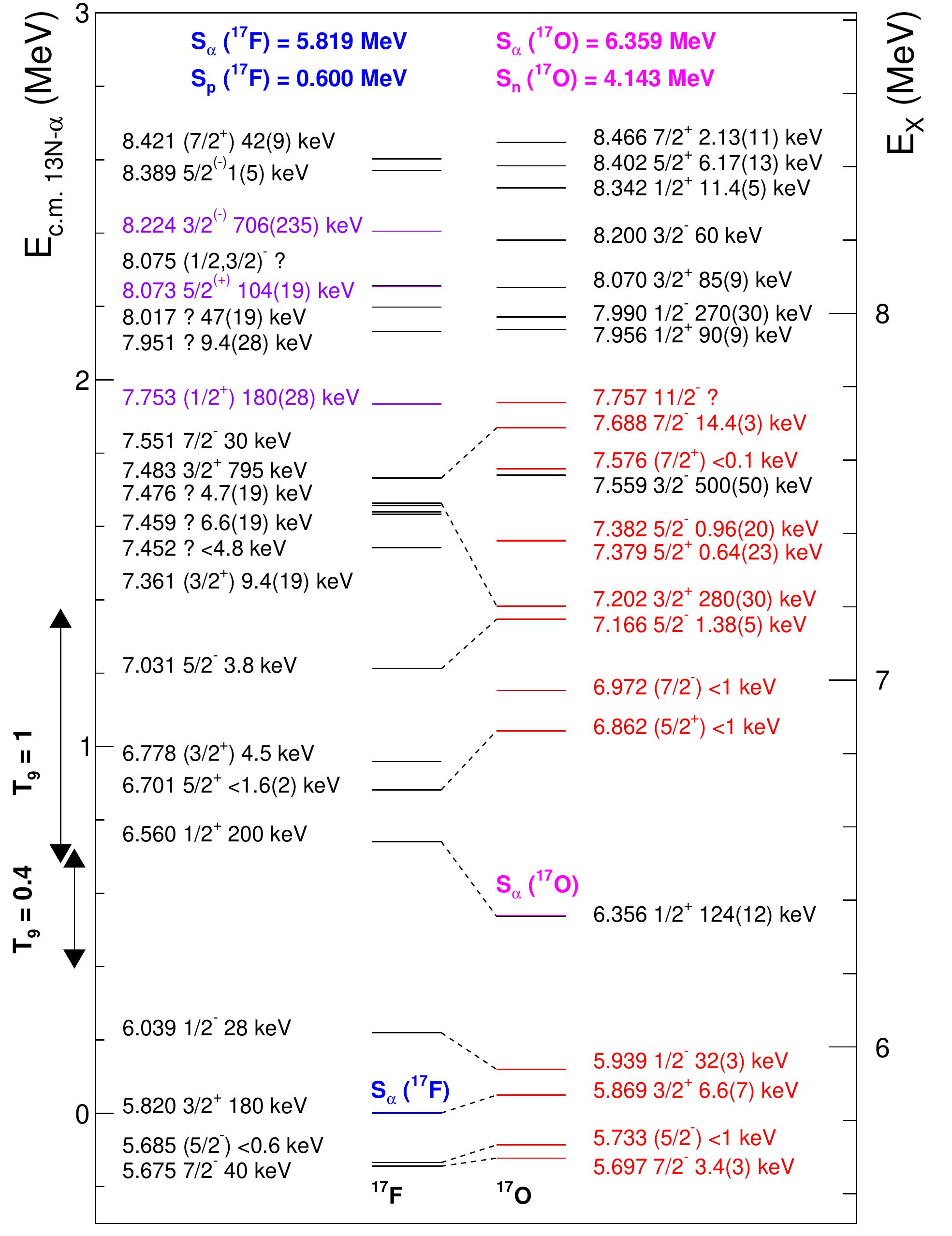}
  \caption{\label{f:mirrors}
     (Color online) Level scheme of \fl\ nucleus above the $\alpha$+$^{13}$N 
     threshold and comparison with its mirror nucleus \ox. Mirror pairs are 
     linked with dashed lines. \ox\ states studied in the present analysis are 
     in red. \fl\ states in purple have experimentally determined partial and 
     total widths. Black arrows indicate the energy range of the Gamow window 
     for two temperatures of interest.}
 \end{figure*}

For temperatures achieved during explosive burning in the He shell of massive 
stars ($T_9 = 0.4 - 1$) the energy range of the Gamow window for the \nap\ 
reaction corresponds to excitation energies of \fl\ between 6.22~MeV and
7.20~MeV. Four \fl\ states are known in this energy region (see 
Fig.~\ref{f:mirrors}), but the tails of broad states lying above the Gamow window could
also contribute to the reaction rate. Hence, in the following, we 
consider \fl\ states having excitation energies up to 8.2~MeV and the relevant 
spectroscopic information is presented in Tab.~\ref{t:mirror}.
                                                                   
States in \fl\ above the $\alpha$+$^{13}$N threshold ($S_{\alpha+\az}$ = 
5818.7~(4)~keV~\cite{Til93}) have mainly been studied by the 
$^{16}$O(p,p)$^{16}$O reaction~\cite{Sal62a,Sal62b} and by the 
$^{16}$O(p,p$^\prime$)$^{16}$O and $^{16}$O(p,$\alpha$)$^{13}$N
reactions~\cite{Dan64}. These experiments measured excitation functions and
were performed by the same group using the University of Wisconsin tandem Van 
de Graaff installation. Spin, parity, total width and energy of the \fl\
states were determined. Energies of the \fl\ states were derived from the 
incident proton beam energy assuming a proton separation energy value 
($S_p$ = 596~keV~\cite{Sal62b,Dan64}) which is now superseded
($S_p$ = 600.27~(25)~keV~\cite{Hua17}). This information was not updated in
the last NNDC compilation~\cite{Til93} but has been taken into account
in Tab.~\ref{t:mirror}. The large reported uncertainty ($\approx20$~keV) 
associated to the energy of most of the \fl\ states (see Tab.~\ref{t:mirror})
comes from a possible error in the calibration of one of the magnets in the beam 
line~\cite{Ajz86}. The excitation energy uncertainty should therefore be better 
considered as a systematic error rather than a statistical uncertainty. 
No uncertainty is reported for the energy of the state at 8.224~MeV although it 
was observed jointly with the states at 7.753~MeV and 8.073~MeV~\cite{Dan64}
for which uncertainties were given. Owing to the large width of the 8.224~MeV 
state ($\Gamma=706$~(235)~keV), and based on the reported energy uncertainties in this excitation energy region~\cite{Til93}, we assign an uncertainty of 40~keV to 
its excitation energy. 

Excitation energies are then used to derive resonance energies using the 
relation $E_r = E_x - S_{\alpha+\az}$, and the uncertainty associated to the 
resonance energy is dominated by the one on excitation energies.

For the \fl\ states under study there is neither experimental determination 
nor theoretical estimate of their partial widths ($\Gamma_p$ and 
$\Gamma_\alpha$), except for the three broad states at 7.753, 8.073 and 
8.224~MeV. The reduced widths ($\gamma^2_i$) of these three broad resonances 
are reported for the $p_0$, $p_1$ and $\alpha_0$ channels~\cite{Dan64} and 
this information was used to calculate the partial widths reported in 
Tab.~\ref{t:mirror}. In the case of the 7.753~MeV state two partial widths sets
are reported~\cite{Dan64}: ($\Gamma_\alpha$, $\Gamma_{p_0}$, $\Gamma_{p_1}$) = 
(11~keV, 135~keV, 34~keV) and (34~keV, 41~keV, 109~keV). Both sets give
similar results for the contribution of the 7.753~MeV state since 
the total width and its energy dependence are very similar in both cases.
We therefore arbitrary choose set~1 (reported in Tab.~\ref{t:mirror}) for 
the partial widths of the 7.753~MeV state.

For the other \fl\ states with no experimental determination of their partial 
widths, they need to be estimated and two different cases are considered 
depending on the existence of a known analog state in \ox.

Pairing of analog states between the \fl\ and \ox\ nuclei was based on their 
spin and parity information and the consistency of their partial and total 
widths. Identified analog states from the present work are connected
by dashed lines in Fig.~\ref{f:mirrors}. For these states we assume that mirror
symmetry holds and that $C^2S_\alpha(\fl) = C^2S_\alpha(\ox)$~\cite{Oli97}. 
The $\alpha$-particle partial width of \fl\ states is then calculated using 
Eq.~\ref{e:width} where the reduced mass and penetrability quantities refer 
to the $\alpha$+$^{13}$N system instead. Note that there are 
some indication of possible charge-symmetry breaking in the lower part of 
the \fl-\ox\ level scheme~\cite{Alb76}.

For \fl\ states with no spectroscopic information and no identified analog
state their $\alpha$-particle partial width must be estimated. In this case
the $\alpha$-width can be calculated using the following formula~\cite{Ili08}
\begin{equation}
  \Gamma_\alpha^{ } = \theta^2_\alpha \times \Gamma^{Wigner}_\alpha, 
  \label{e:wigner}
\end{equation}
where $\theta^2_\alpha$ is the dimensionless reduced $\alpha$-width and 
$\Gamma^{Wigner}_\alpha = 2 \hbar^2/(\mu r^2) \times P_L(r,E)$ is the Wigner 
limit. We used a mean reduced alpha-width of $\langle\theta^2_\alpha\rangle = 
0.04$ following the same approach as in Ref.~\cite{Moh14}. This value was
obtained from an extrapolation of a data set providing mean dimensionless
$\alpha$-particle reduced widths from nuclei having slightly larger mass 
numbers $A$~\cite{Pog13}. 

For all determinations of the \fl\ $\alpha$-particle partial widths in the 
present work we use the same channel radius $r=7.5$~fm as for the determination 
of $\Gamma_\alpha(\ox)$, which corresponds to $r_0 = 1.9$~fm where $r_0$ is 
defined as $r=r_0\times(A_\alpha^{1/3} + A_{^{13}\mathrm{N}}^{1/3})$. Proton 
widths are deduced in all cases as $\Gamma_p=\Gamma_{tot}-\Gamma_\alpha$, 
except in the case of the three broad states at 7.753, 8.073 and 8.224~MeV. 
The \fl\ resonance parameters derived from this work are summarized in 
Tab.~\ref{t:mirror}, and spectroscopic information of \ox\ states is given 
when pairing of analog states is established.

 \begin{table*}[!htpb] 
  \caption{\label{t:mirror}
    Resonance parameters in \fl\ above $^{13}$N+$\alpha$ threshold ($S_{\alpha+\az}$ 
    = 5818.7 (4)~keV) and spectroscopic information for the \ox\ analog states 
    when available. \ox\ state properties come from NNDC~\cite{Til93} unless 
    otherwise stated.}
  \begin{ruledtabular}
  \begin{tabular}{ccccccccccccc}
    \multicolumn{8}{c}{$^{17}$F} & \multicolumn{5}{c}{$^{17}$O} \\ \cline{1-8} \cline{9-13}
    $E_x$\footnotemark[1] & $E_r$ & $J^\pi$ & $\ell_\alpha$, $\ell_p$ & $\Gamma_\alpha$\footnotemark[2] & $\Gamma_{p_0}$\footnotemark[3] & $\Gamma_{p_1}$ & $\Gamma_{tot}$\footnotemark[4] &
    $E_x$ & $J^\pi$  & $\Gamma_\alpha$ & $\Gamma_n$ & $\Gamma_{tot}$ \\         
    (MeV) & (keV)    &               &        & (keV) & (keV)         & (keV) & (keV) & (MeV) &         & (keV)          & (keV)      & (keV) \\ \hline
    5.820 (20) &  1.3 &  3/2$^+$  & 1, 2 & 6.92$\times10^{-283}$\footnotemark[8] & 180 & & 180 &                                     5.869 &  3/2$^+$  & & 6.6   & 6.6 (7)              \\
    6.039  (9) &  221 &  1/2$^-$  & 0, 1 & 2.63$\times10^{-13}$ & 28  & &  28 &            5.939 &  1/2$^-$  & & 31.5  & 32 (3) \\
    6.560 (20) &  741 &  1/2$^+$  & 1, 0 & 1.88$\times10^{-3}$  & 200 & & 200 &            6.356 &  1/2$^+$  & & 124   & 124 (12) \\
    6.701  (7) &  882 &  5/2$^+$  & 3, 2 & 1.76$\times10^{-5}$  & 1.6 & & $\leq$ 1.6 (2) & 6.862 & (5/2$^+$) & & & $<$ 1 \\
    6.778 (20) &  959 & (3/2$^+$) & 1, 2 & 3.00$\times10^{-2}$  & 4.47 & & 4.5 &                                           &         & & &   \\
    7.031 (20) & 1213 &  5/2$^-$  & 2, 3 & 3.59$\times10^{-2}$  & 3.76 & & 3.8 &           7.166 &  5/2$^-$  & 0.0033 & 1.38 (5) & 1.38 (5) \\
    7.361 (20) & 1542 & (3/2$^+$) & 1, 2 & 2.20 & 7.20 & & 9.4 (19) &                            &           & & &                                 \\
    7.452 (20) & 1633 &           & & & & & $\leq$ 4.7 &                                         &           & & &                                 \\
    7.459 (20) & 1640 &           & & & & & 6.6 (19) &                                           &           & & &                                 \\
    7.476 (20) & 1657 &           & & & & & 4.7 (19) &                                           &           & & &                                 \\
    7.483 (20) & 1664 &  3/2$^+$  & 1, 2 & 4.64 & 790.36 & & 795 &                         7.202 &  3/2$^+$  & 0.07 & 280 & 280 (30)          \\
    7.551 (20) & 1732 &  7/2$^-$  & 4, 3 & 1.10$\times10^{-2}$ & 29.98 & & 30 &            7.688 &  7/2$^-$  & 0.01 & 13.0 (6) & 14.4 (3)     \\
    7.753 (40) & 1935 &  (1/2$^+$)\footnotemark[5]  & 1, 0 & 11\footnotemark[6] & 135\footnotemark[6] & 34\footnotemark[6] & 180 (28) & 7.956 
    &  1/2$^+$  & 6.7 & 84 & 90 (9)  \\
    7.951 (30) & 2132 &           & & & & & 9.4 (28) &                                           &           & & &                                 \\
    8.017 (40) & 2198 &           & & & & & 47 (19) &                                            &           & & &                                 \\
    8.073 (30) & 2255 & 5/2$^{(+)}$\footnotemark[5]  & 3, 2 & 14\footnotemark[6] & 79\footnotemark[6] & 11\footnotemark[6] & 104 (19) &          &           & & &                                 \\
    8.075 (10) & 2256 & (1/2,3/2)$^-$ & 0-2, 1 &  &  &  &  &                                    &           & & &                                 \\
    8.224 (40)\footnotemark[7] & 2405 & 3/2$^{(-)}$\footnotemark[5] & 2, 1 & 25\footnotemark[6] & 636\footnotemark[6] & 45\footnotemark[6] & 706 (235) &         &           & & &                          \\
  \end{tabular}                                                                
  \end{ruledtabular}
    \footnotetext[1] {Energies have been corrected when needed with the new $^{16}$O+p threshold value ($S_p=600.27$~keV~\cite{Hua17}), see text. 
      Uncertainties are from the latest compilation~\cite{Til93}. Note that reported uncertainties
      greater than 10~keV used to be smaller by a factor of two (see footnote $a$ in Table~17.19~\cite{Ajz86}.)}
    \footnotetext[2] {When a mirror connection exists the same reduced width 
      $\gamma^2_\alpha$ is assumed between analog states. Otherwise a 
      dimensionless reduced width $\langle\theta^2_\alpha\rangle = 0.04$ is 
      assumed~\cite{Pog13,Moh14}. In all cases a channel radius of 7.5~fm is 
      used.}
    \footnotetext[3] {$\Gamma_{p_0} = \Gamma_{tot} - \Gamma_\alpha$}
    \footnotetext[4] {Total widths have been transformed to center of mass values when needed.}
    \footnotetext[5] {While parity for these three states is uncertain, their relative ordering is fixed~\cite{Dan64}.}
    \footnotetext[6] {$\Gamma_{p_0}$, $\Gamma_{p_1}$ and $\Gamma_\alpha$ are 
      deduced from reduced widths derived from $^{16}$O(p,p)$^{16}$O~\cite{Sal62a,Sal62b} and        
      $^{16}$O(p,p$^\prime$)$^{16}$O and $^{16}$O(p,$\alpha$)$^{13}$N~\cite{Dan64} measurements.}
    \footnotetext[7] {Uncertainty is set arbitrarily from present work (see 
      text).}
    \footnotetext[8] {Despite an established mirror connection, a dimensionless reduced width $\langle\theta^2_\alpha\rangle = 0.04$ is assumed since the alpha spectroscopic factor for the $E_x = 5.869$~MeV state in \ox\ could not be determined (see text).
    }
 \end{table*}
 
\begin{figure}[!htpb]
  \includegraphics[width=0.53\textwidth]{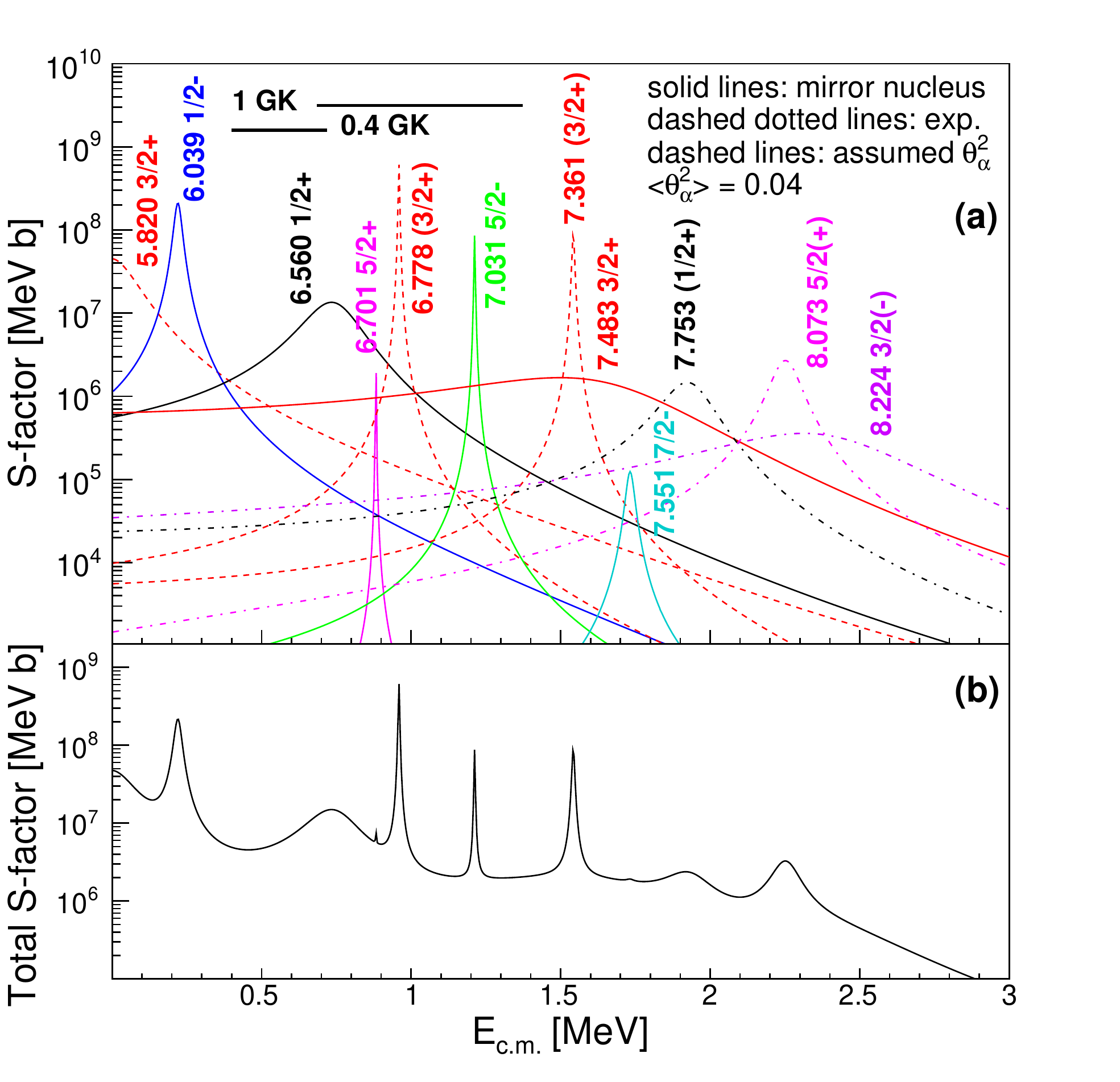}
  \caption{\label{f:sfact}
    (Color online) Astrophysical S-factor for the \nap\ reaction as a function
    of the center-of-mass energy. 
    $\mathcal{R}$-matrix calculations using the AZURE2 code with the 
    parameters given in Tab.~\ref{t:mirror} are represented for individual 
    resonances in panel (a). The Gamow energy window is represented for the temperatures $T_9=0.4$ and $T_9=1$. Panel (b) represents the total astrophysical S-factor when individual contributions from panel (a) are summed.}
\end{figure}

The contribution of individual \fl\ resonances to the astrophysical S-factor 
$S(E)$ of the \nap\ reaction, calculated using the spectroscopic information
given in Tab.~\ref{t:mirror}, is shown in Fig.~\ref{f:sfact}a. Calculations 
were performed with the $\mathcal{R}$-matrix code AZURE2~\cite{Azu10} using
channel radius $r_\alpha=7.5$~fm and $r_p=6.7$~fm. Solid lines correspond to
resonances for which the $\alpha$-particle partial width is estimated from the
analog states, and dashed lines correspond to resonances where 
$\langle\theta^2_\alpha\rangle=0.04$ is assumed. The major contribution to
the S-factor in the temperature range of interest comes from the broad 
$E_r=741$~keV and the two narrow $E_r=959$ and 1213~keV resonances 
corresponding to low $\ell_\alpha$ angular momentum. Resonances lying outside 
the Gamow window have a minor contribution in the energy region of interest,
except in case of the broad $E_r=1664$~keV resonance for the highest 
temperatures ($T_9=1-2$). The total astrophysical S-factor obtained when all
individual contributions are summed is shown in Fig.~\ref{f:sfact}b.

\section{MONTE-CARLO REACTION RATES} \label{rates}
\subsection{Method}
The reaction rate per particle pair is defined as~\cite{Ili08}
\begin{equation}
  \langle\sigma v\rangle                                                        
  = \left( \frac{8}{\pi\mu} \right)^{1/2}   \frac{1}{(kT)^{3/2}}         
  \int_{0}^{\infty} E \sigma(E) e^{-E/kT} dE          
  \label{e:rate}
\end{equation}
where $\mu$ is the reduced mass of the interacting particles, $k$ is the 
Maxwell-Boltzmann constant, $T$ is the temperature, and $\sigma(E)$ is 
the nuclear reaction cross section. In the present case the \nap\ 
reaction proceeds through several resonances and the cross section 
associated to a single resonance is defined by the one-level Breit-Wigner 
formula
\begin{equation}
  \sigma(E) = \frac{\lambda^2}{4\pi} \frac{(2J+1)}{4} 
              \frac{\Gamma_\alpha(E)\Gamma_p(E+Q)}{(E - E_r)^2 + \Gamma/4} 
  \label{e:xsec}
\end{equation}
where $\lambda$ is the de Broglie wavelength, $J$ and $E_r$ are the spin and
energy of the \fl\ resonance, respectively, $\Gamma_i$ are the energy dependent
partial widths and $\Gamma$ is the total width.

In order to determine a statistically meaningful \nap\ reaction rate the Monte 
Carlo method developed by Ref.~\cite{Lon10} has been followed. In summary, the 
energy and partial widths of each resonance are varied according to the 
probability density function defined by the experimental mean value and the 
associated uncertainty. For a given variation of the resonance energy, the 
partial widths are consistently evaluated by using the correct energy in the
determination of the penetrability of the Coulomb and centrifugal barriers.
For each Monte Carlo realization, all uncertain resonance parameters are 
sampled and a reaction rate is calculated. For a sufficiently large number of 
realizations (10000 in the present work), a statistical meaningful recommended,
low and high reaction rates can be defined. They are defined in this work as 
the 50$^{th}$, 16$^{th}$ and 84$^{th}$ percentile of the cumulative rate 
distribution, respectively.

Two different probability density functions are used for sampling the 
alpha-width of \fl\ states depending on whether or not an analog \ox\ state 
is known. When this is known, a lognormal distribution is used and an 
uncertainty of a factor of 2.5 on the alpha particle width is assumed. This 
uncertainty comes from the combination of the uncertainty on the \ox\ alpha 
spectroscopic factor deduced from the transfer reaction ($\approx50\%$) and 
the assumption of mirror symmetry which accounts for a factor of two 
uncertainty when states with relatively large spectroscopic factors are 
considered~\cite{Moh14}. In case of \fl\ states with no identified analog state, 
the $\alpha$-width is sampled according to a Porter-Thomas distribution of 
dimensionless reduced alpha-width $\langle\theta^2_\alpha\rangle = 
0.04\pm0.02$. 
 
Concerning the proton width of \fl\ states a lognormal distribution is used 
and an uncertainty of 20\% is assumed when no such uncertainty is reported 
in the literature. In the case of the three broad states measured directly we 
estimate an uncertainty for their alpha and proton widths assuming the same 
relative uncertainty as for their total width.

For the resonance energies we assume a Gaussian probability density 
function. Usually, energy uncertainties are considered independent from each 
others, which is a valid assumption when energy determination comes from 
different experimental techniques where systematic uncertainties are expected
to be uncorrelated. In the present case, however, all states having energy 
uncertainties greater than 20~keV have been studied by the same group at the 
same facility using the same experimental technique, which leads to highly 
correlated uncertainties (see Section~\ref{fluor17}). Here we extend the Monte 
Carlo method by implementing correlated energy uncertainties for several 
resonances following a similar approach as for the correlated uncertainties 
on resonance strengths~\cite{Lon17}. First the smallest energy uncertainty is 
identified (20~keV in the present case), then the ratio of this value to each individual resonance energy uncertainty, $\sigma_{Ej}$, is used to calculate a correlation factor, $\rho_j$. Two cases are considered: (i) a resonance with an uncertainty equal to the 20-keV minimum uncertainty in the present case. For this resonance, $\rho=1$; and (ii) a resonance with a much larger uncertainty, say 40~keV, yielding $\rho=20/40=0.5$. During the Monte Carlo procedure, each resonance energy sample, $E_{j,i}$, for resonance $j$ is computed using the following procedure. A reference sample, $x_{r,i}$, and uncorrelated samples for each resonance, $y_{j,i}$, are obtained from a Normal distribution (that is, a Gaussian distribution with a mean, $\mu=0$, and standard deviation, $\sigma = 1$). Correlated, normally distributed random samples for each resonance are then calculated using:
\begin{equation}
    y'_{j,i} = \rho_j x_{r,i} + \sqrt{1-\rho_j^2} y_{j,i}.
\end{equation}
Finally, the resonance energy samples are calculated using
\begin{equation}
    E_{j,i} = E_j + \sigma_{Ej} y'_{j,i}.
\end{equation}

For \fl\ states where spin and parity assignments are uncertain, a range of 
possible $J^\pi$ defined by $\ell_\alpha\pm1$ is considered, where $\ell_\alpha$
is the tentative alpha orbital angular momentum given in Tab.~\ref{t:mirror}. 
This range is then sampled according to a discrete probability density function 
for each Monte Carlo realization. Following the approach of Mohr et al.~\cite{Moh14} a probability of 50\% is taken for the 
tentative spin and  parity while the remaining 50\% are equally shared between the 
other spin and parity possibilities.

\subsection{\nap\ reaction rate}
\begin{figure}[!htpb]
  \includegraphics[width=0.5\textwidth]{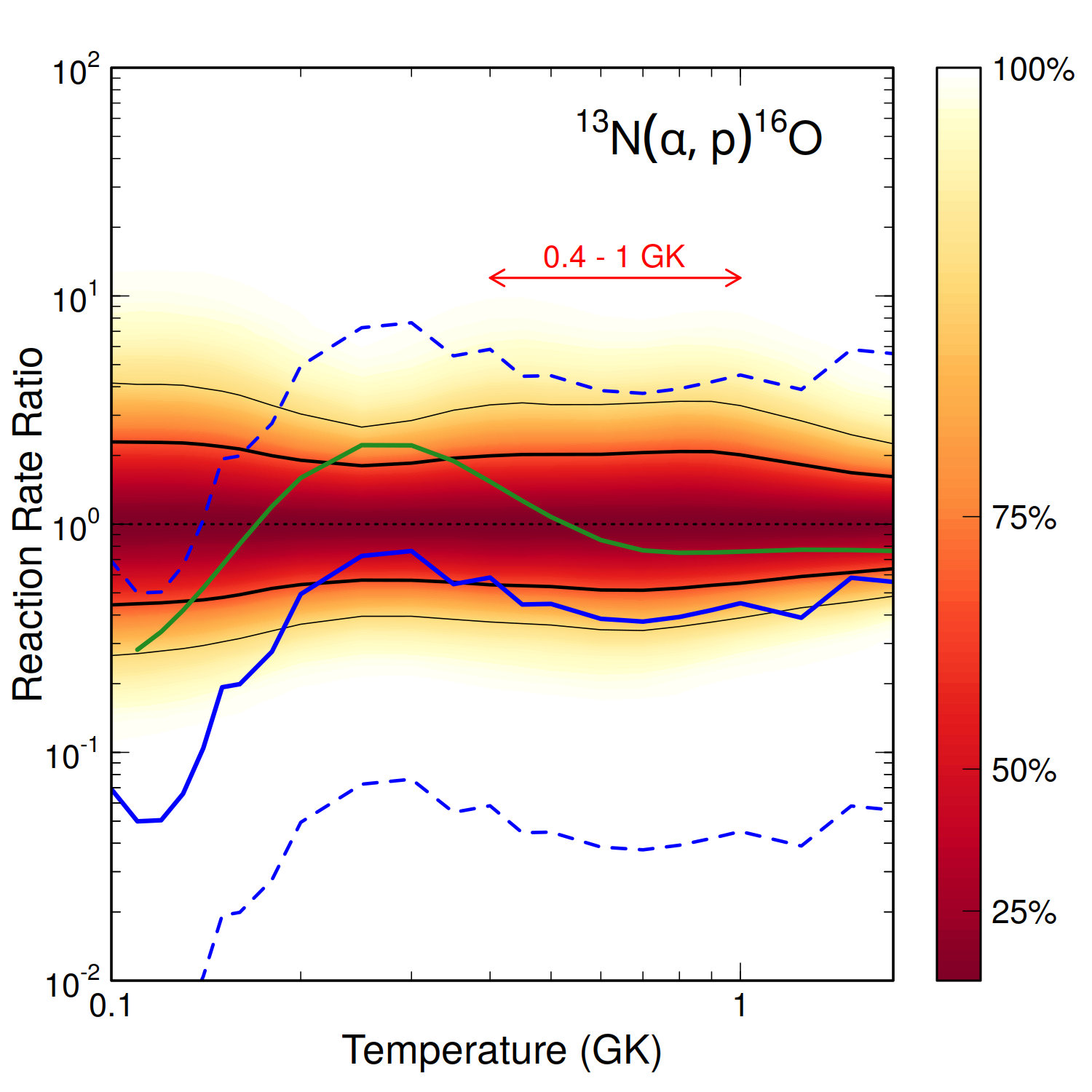}
  \caption{\label{f:rate}
    (Color online) Ratio of different \nap\ reaction rates normalized to the
    recommended reaction rate defined as the 50$^{th}$ percentile of the 
    cumulative rate distribution obtained from the Monte Carlo procedure.
    The area delimited by the thick/thin black lines comprise a
    coverage probability of 68\%/95\%, respectively. The green line corresponds
    to the \nap\ reaction rate given by CF88, while the blue lines represent
    the nominal \nap\ reaction rate with associated uncertainty from STARLIB.}
\end{figure}

The results of the Monte Carlo simulation for the \nap\ reaction rates are
presented in Fig.~\ref{f:rate}, where all rates are normalized to the 
recommended reaction rate defined in the previous section. 
The colored area represents a coverage probability of 68\% which 
corresponds to an uncertainty of a factor of about two to three at 
the temperature of interest $T_9 = 0.4 - 1$. This is not surprising since the reaction
rate in this temperature range is dominated by the contribution of the 221- 
and 741-keV resonances for which the alpha-widths are determined from the known \ox\ 
analog states with a factor uncertainty of 2.5. The \nap\ reaction rate
from Caughlan \& Fowler~\cite{CF88} is represented by the green curve and is
within a factor of three of the recommended rate across all the temperature 
range and within less than a factor of two between $T_9 = 0.4 - 1$. The \nap\ 
reaction rate from the STARLIB library~\cite{Sal13} based on Hauser-Feschbach 
theory is represented as the blue curve. The temperature dependence is somewhat 
similar to the Caughlan \& Fowler rate, but the STARLIB rate is systematically 
lower. For the temperature range of interest, $T_9 = 0.4 - 1$, the STARLIB rate 
is lower than the recommended \nap\ reaction rate from the present work by a factor two.

\begin{figure}[!htpb]
  \includegraphics[width=0.5\textwidth]{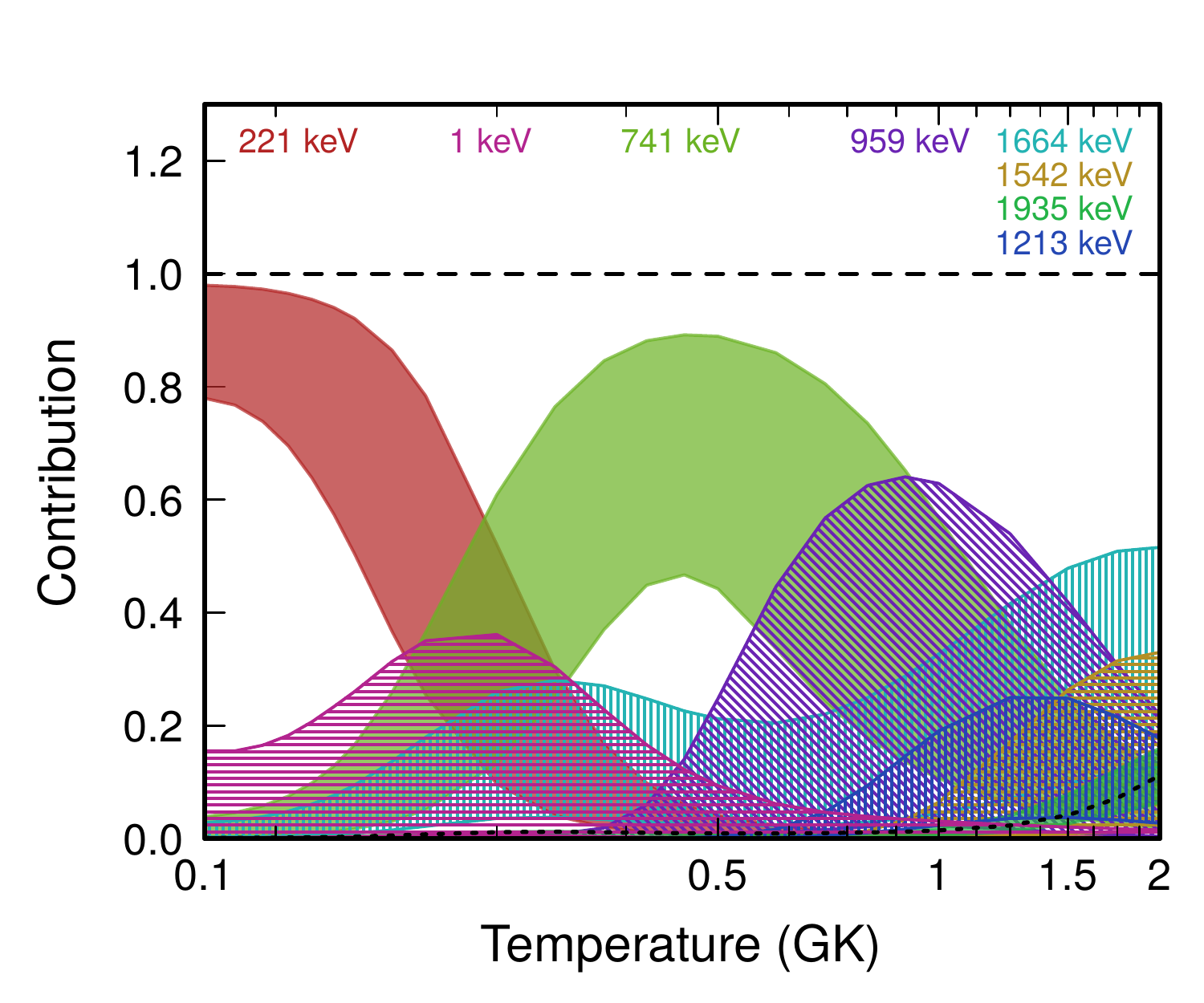}
  \caption{\label{f:contrib}
    (Color online) Fractional contribution of individual resonances to the 
    \nap\ reaction rate. The numbers at the top of the Figure correspond to
    the center of mass energy of each resonance.}
\end{figure}

The fractional contribution of individual resonances to the \nap\ reaction
rate is represented in Fig.~\ref{f:contrib}. Three resonances at $E_r^{c.m.}$ = 
221, 741 and 959~keV are dominating the \nap\ reaction rate, the latter being
the major contributor in the temperature range of interest $T_9 = 0.4 - 1$.
While at $T_9 = 0.4$ the \nap\ reaction rate is mostly dominated by the 
single resonance at 741~keV, several resonances contribute at $T_9 = 1$. 
The case of the $E_r^{c.m.}$ = 959~keV resonance is interesting since its
relative contribution can be consistent with zero or as high as 60\% at $T_9 = 1$.
The broad resonance at 1664~keV may contribute across all the temperature range of interest because of its large natural width ($\Gamma = 795$~keV).

In this work resonances up to an energy of 2.4~MeV are considered. This 
corresponds to a cutoff temperature of 1.4~GK when the procedure relying on the
cumulative distribution of fractional resonant rates given in~Ref.~\cite{New08} 
is followed. Below this temperature the low, recommended and high \nap\ reaction
rates come from the present Monte Carlo study. At higher temperatures the 
recommended reaction rate is calculated by normalizing the \nap\ Hauser-Feshbach 
reaction rate given in the STARLIB database~\cite{Sal13}. The 
reaction rates are given numerically in Tab.~\ref{t:rates}.

\begin{table}[!htpb]
  \caption{\label{t:rates}
     Low, recommended and high thermonuclear rates of the \nap\ reaction are given in 
     cm$^3$ s$^{-1}$ mol$^{-1}$ as a function of temperature ($T_9$). Rates are derived 
     from a Monte Carlo approach 
     below the cutoff temperature ($T_9=1.4$) (see text), and
     come from the STARLIB database at higher temperatures.}
  \begin{ruledtabular}
  \begin{tabular}{lccc}
    $T_9$ & Low & Recommended & High \\ \hline
0.01	&	8.63$\times$10$^{-55}$	&	3.07$\times$10$^{-54}$	&	1.25$\times$10$^{-53}$ \\
0.011	&	1.48$\times$10$^{-52}$	&	5.15$\times$10$^{-52}$	&	2.06$\times$10$^{-51}$ \\
0.012	&	1.41$\times$10$^{-50}$	&	4.77$\times$10$^{-50}$	&	1.89$\times$10$^{-49}$ \\
0.013	&	8.25$\times$10$^{-49}$	&	2.73$\times$10$^{-48}$	&	1.07$\times$10$^{-47}$ \\
0.014	&	3.23$\times$10$^{-47}$	&	1.06$\times$10$^{-46}$	&	4.10$\times$10$^{-46}$ \\
0.015	&	9.09$\times$10$^{-46}$	&	2.93$\times$10$^{-45}$	&	1.12$\times$10$^{-44}$ \\
0.016	&	1.93$\times$10$^{-44}$	&	6.13$\times$10$^{-44}$	&	2.30$\times$10$^{-43}$ \\
0.018	&	4.32$\times$10$^{-42}$	&	1.33$\times$10$^{-41}$	&	4.81$\times$10$^{-41}$ \\
0.02	&	4.57$\times$10$^{-40}$	&	1.36$\times$10$^{-39}$	&	4.78$\times$10$^{-39}$ \\
0.025	&	5.25$\times$10$^{-36}$	&	1.44$\times$10$^{-35}$	&	4.77$\times$10$^{-35}$ \\
0.03	&	6.77$\times$10$^{-33}$	&	1.74$\times$10$^{-32}$	&	5.31$\times$10$^{-32}$ \\
0.04	&	2.46$\times$10$^{-28}$	&	5.85$\times$10$^{-28}$	&	1.54$\times$10$^{-27}$ \\
0.05	&	5.16$\times$10$^{-25}$	&	1.16$\times$10$^{-24}$	&	2.70$\times$10$^{-24}$ \\
0.06	&	2.23$\times$10$^{-22}$	&	4.99$\times$10$^{-22}$	&	1.12$\times$10$^{-21}$ \\
0.07	&	3.16$\times$10$^{-20}$	&	7.05$\times$10$^{-20}$	&	1.60$\times$10$^{-19}$ \\
0.08	&	1.73$\times$10$^{-18}$	&	3.89$\times$10$^{-18}$	&	8.83$\times$10$^{-18}$ \\
0.09	&	4.43$\times$10$^{-17}$	&	1.00$\times$10$^{-16}$	&	2.29$\times$10$^{-16}$ \\
0.1	&	6.35$\times$10$^{-16}$	&	1.44$\times$10$^{-15}$	&	3.29$\times$10$^{-15}$ \\
0.11	&	5.92$\times$10$^{-15}$	&	1.32$\times$10$^{-14}$	&	3.02$\times$10$^{-14}$ \\
0.12	&	3.93$\times$10$^{-14}$	&	8.69$\times$10$^{-14}$	&	1.98$\times$10$^{-13}$ \\
0.13	&	2.03$\times$10$^{-13}$	&	4.42$\times$10$^{-13}$	&	1.00$\times$10$^{-12}$ \\
0.14	&	8.66$\times$10$^{-13}$	&	1.86$\times$10$^{-12}$	&	4.15$\times$10$^{-12}$ \\
0.15	&	3.19$\times$10$^{-12}$	&	6.69$\times$10$^{-12}$	&	1.46$\times$10$^{-11}$ \\
0.16	&	1.05$\times$10$^{-11}$	&	2.13$\times$10$^{-11}$	&	4.56$\times$10$^{-11}$ \\
0.18	&	8.73$\times$10$^{-11}$	&	1.67$\times$10$^{-10}$	&	3.33$\times$10$^{-10}$ \\
0.2	&	5.52$\times$10$^{-10}$	&	1.01$\times$10$^{-09}$	&	1.93$\times$10$^{-09}$ \\
0.25	&	2.70$\times$10$^{-08}$	&	4.75$\times$10$^{-08}$	&	8.57$\times$10$^{-08}$ \\
0.3	&	6.43$\times$10$^{-07}$	&	1.13$\times$10$^{-06}$	&	2.09$\times$10$^{-06}$ \\
0.35	&	9.13$\times$10$^{-06}$	&	1.64$\times$10$^{-05}$	&	3.19$\times$10$^{-05}$ \\
0.4	&	8.64$\times$10$^{-05}$	&	1.59$\times$10$^{-04}$	&	3.17$\times$10$^{-04}$ \\
0.45	&	5.88$\times$10$^{-04}$	&	1.10$\times$10$^{-03}$	&	2.21$\times$10$^{-03}$ \\
0.5	&	3.03$\times$10$^{-03}$	&	5.70$\times$10$^{-03}$	&	1.15$\times$10$^{-02}$ \\
0.6	&	4.20$\times$10$^{-02}$	&	8.15$\times$10$^{-02}$	&	1.65$\times$10$^{-01}$ \\
0.7	&	3.15$\times$10$^{-01}$	&	6.14$\times$10$^{-01}$	&	1.26$\times$10$^{+00}$ \\
0.8	&	1.57$\times$10$^{+00}$	&	3.00$\times$10$^{+00}$	&	6.24$\times$10$^{+00}$ \\
0.9	&	5.94$\times$10$^{+00}$	&	1.10$\times$10$^{+01}$	&	2.29$\times$10$^{+01}$ \\
1	&	1.82$\times$10$^{+01}$	&	3.30$\times$10$^{+01}$	&	6.65$\times$10$^{+01}$ \\
1.25	&	1.67$\times$10$^{+02}$	&	2.84$\times$10$^{+02}$	&	5.17$\times$10$^{+02}$ \\
1.5	&	8.66$\times$10$^{+02}$	&	1.41$\times$10$^{+03}$	&	2.37$\times$10$^{+03}$ \\
1.75	&	2.88$\times$10$^{+03}$	&	4.68$\times$10$^{+03}$	&	7.86$\times$10$^{+03}$ \\
2	&	9.55$\times$10$^{+03}$	&	1.55$\times$10$^{+04}$	&	2.61$\times$10$^{+04}$ \\
2.5	&	4.88$\times$10$^{+04}$	&	7.95$\times$10$^{+04}$	&	1.34$\times$10$^{+05}$ \\
3	&	1.62$\times$10$^{+05}$	&	2.63$\times$10$^{+05}$	&	4.42$\times$10$^{+05}$ \\
3.5	&	4.06$\times$10$^{+05}$	&	6.61$\times$10$^{+05}$	&	1.11$\times$10$^{+06}$ \\
4	&	8.48$\times$10$^{+05}$	&	1.38$\times$10$^{+06}$	&	2.32$\times$10$^{+06}$ \\
5	&	2.57$\times$10$^{+06}$	&	4.19$\times$10$^{+06}$	&	7.04$\times$10$^{+06}$ \\
6	&	5.76$\times$10$^{+06}$	&	9.37$\times$10$^{+06}$	&	1.57$\times$10$^{+07}$ \\
7	&	1.06$\times$10$^{+07}$	&	1.73$\times$10$^{+07}$	&	2.91$\times$10$^{+07}$ \\
8	&	1.72$\times$10$^{+07}$	&	2.80$\times$10$^{+07}$	&	4.71$\times$10$^{+07}$ \\
9	&	2.54$\times$10$^{+07}$	&	4.13$\times$10$^{+07}$	&	6.95$\times$10$^{+07}$ \\
10	&	3.49$\times$10$^{+07}$	&	5.68$\times$10$^{+07}$	&	9.55$\times$10$^{+07}$ \\
  \end{tabular}                                                                
  \end{ruledtabular}
\end{table}

\subsection{Discussion}
The main source of uncertainty for the \nap\ reaction rate comes from the 2.5 
factor associated to the alpha-widths uncertainty for resonances having a 
known \ox\ analog state. This is particularly true for the $E_r^{c.m.}$ = 221~keV ($E_x=6.039$~MeV) 
and 741~keV ($E_x=6.560$~MeV) resonances in the $T_9 = 0.4 - 1$ range. Reducing these
uncertainties should be the first priority for future dedicated experimental 
work. The remaining uncertainty are caused by the unknown spins and parities
together with the large correlated energy uncertainty.
Additional Monte Carlo reaction rate calculations have been performed assuming smaller
uncertainties for the spectroscopic properties (spin/parity, energy, partial widths) of the
$\alpha+^{13}$N resonances. These calculations show a reduction of the uncertainty on the \nap\ reaction rate
but the recommended rate does not vary by more than 10\%.
Similarly, the effect of the uncertainty on the $\theta^2_\alpha$ parameter
has been investigated considering two additional cases, e.g. 
$\theta^2_\alpha=0.03\pm0.02$ and $\theta^2_\alpha=0.05\pm0.02$. As in 
Ref.~\cite{Moh14} we find that the uncertainty on this value has a minor impact on the final recommended \nap\ reaction rate.

Interference effects have been neglected in this work given the current level 
of uncertainty on the spin and parity, and the resonance strengths, of states 
within 2.4~MeV above the $^{13}$N+$\alpha$ threshold. The level at 6.560~MeV 
could interfere with the level at 7.753~MeV if its spin-parity assignment (1/2$^+$) is 
confirmed. However, the effect of either constructive or destructive 
interferences would be hindered by the contribution of the broad 7.483~MeV 
state. The case of interfering 3/2$^+$ states is different since the broad 
7.483~MeV state ($\Gamma_{tot}=795$~keV) can interfere with the two potential 
3/2$^+$ states at 6.778 and 7.361~MeV. The impact of these interferences would 
be most noticeable between the two levels at 6.778 and 7.361~MeV, well within 
the Gamow energy window for $T_9=1$. At lower energies, below the 6.778~MeV 
state, interference effects would be obscured by the 6.560~MeV contribution.
Reaction rate calculations of the cases discussed above have shown that the interference effects account for at most a few percent change in the recommended reaction rate.

The contribution to the reaction rate of the states at $E_x=7.452, 7.459, 
7.476, 7.951$ and 8.017~MeV has not been taken into account since their spins
and parities are not known. However their impact has been estimated assuming
these states have $J^\pi=1/2^-$ ($\ell_\alpha=0$) and a dimensionless reduced
alpha-width $\theta^2_\alpha=0.04$. $\mathcal{R}$-matrix calculations show
that none of these resonances can contribute significantly for $T_9\leq1$, and
therefore they can be safely neglected in this temperature regime. This
situation arises from the rather small total width of these resonances
($\sim5-50$~keV) located at energies well above the upper bound of the Gamow
peak for $T_9=1$ ($E=1.375$~MeV).

\section{ASTROPHYSICAL IMPLICATIONS} 
\label{astro}
To understand the impact of the new rate of the \nap~reaction, we have performed single-zone post-processing nucleosynthesis simulations.
Sixteen explosive trajectories including temperature and densities evolving over time were extracted from the He shell of the 15~$M_\odot$, metallicity (Z) = 0.02 core-collapse supernova (CCSN)
model by \cite{fryer:18} (Samuel Jones and Chris Fryer, private communication). These trajectories are representative of a range of 0.4 GK $\lesssim$ $T$ $\lesssim$ 0.7 GK for the peak temperature at the passage of the SN shock. For the initial abundances, we used the He shell pre-explosive composition between mass coordinates 6.95~$M_\odot$ and 7.05~$M_\odot$, from the 25~$M_\odot$, Z = 0.02~massive star model by \cite{pignatari:16}, following the same approach used by \cite{pignatari:15}. In particular, it is relevant to use this initial composition since the 25~$M_\odot$ stellar model experienced H ingestion in the He shell, and therefore its abundance signature will be representative for the impact study provided in this work.
The He-rich shell material is left with about 1.2~\% of H.

\begin{figure}[!hbtp]
  \begin{minipage}[b]{0.5\textwidth}
    \hspace{-0.5cm}
    \includegraphics[width=0.98\textwidth]{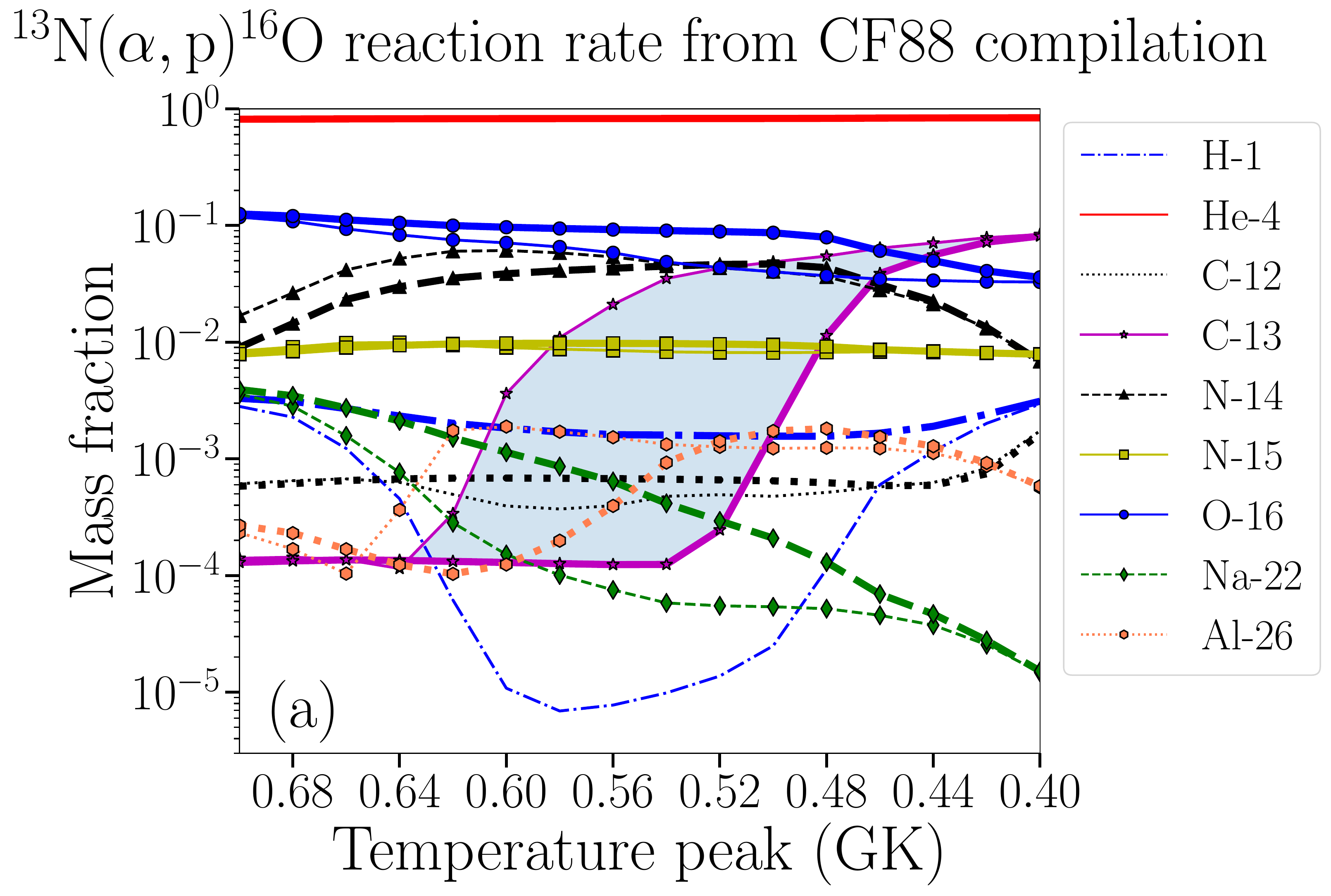}
    \label{f:massFracReaclib}
  \end{minipage}
  \begin{minipage}[b]{0.5\textwidth}
    \includegraphics[width=0.94\textwidth]{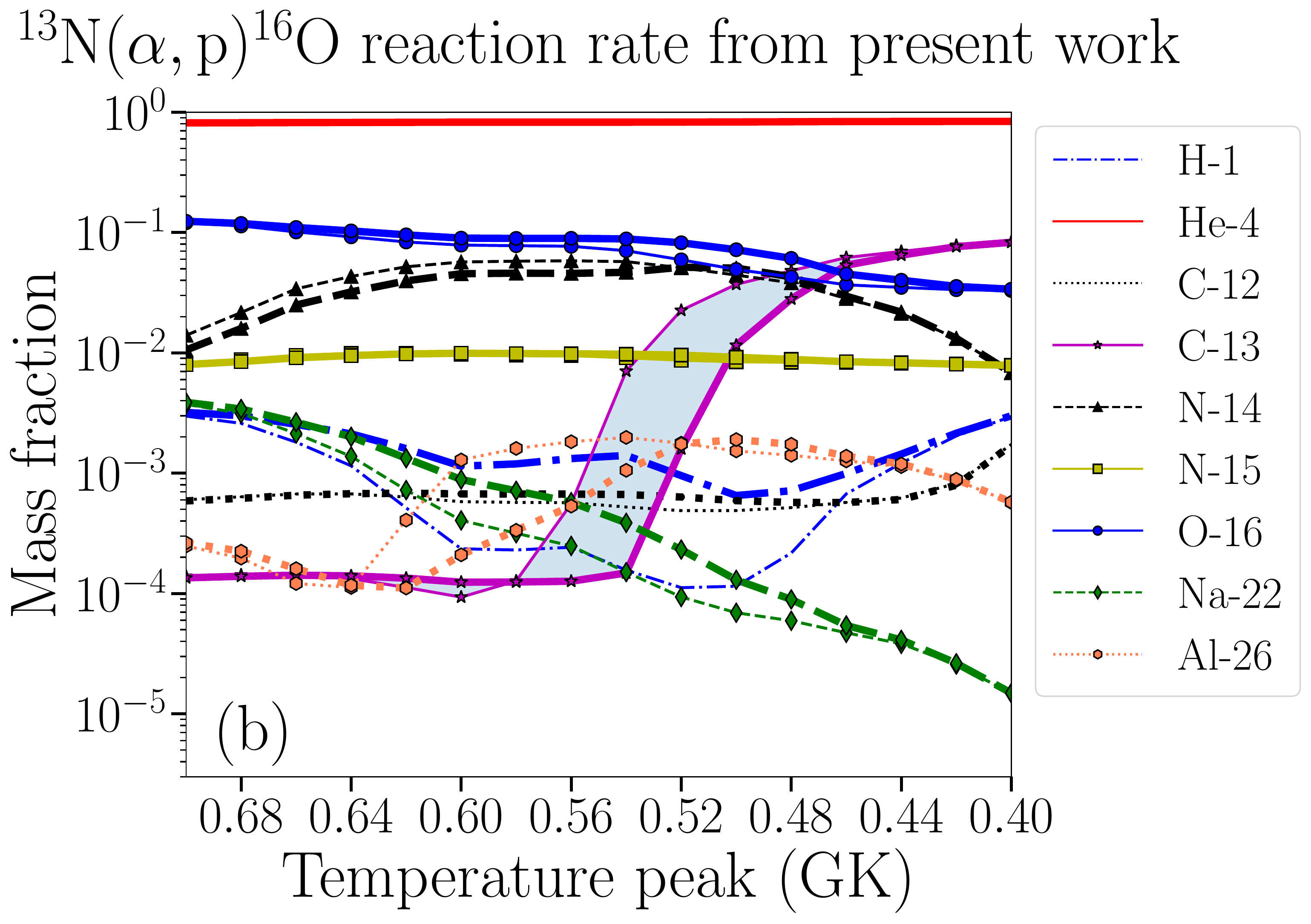}
    \label{f:massFracNewRate}
  \end{minipage}
  \caption{\label{f:massFrac} 
     (Color online) Isotopic abundances in the He-shell ejecta of a 25~$M_\odot$ supernova model. 
     Upper panel: impact of a variation of the \nap~reaction rate by an arbitrary factor of five with
     respect to the CF88 rate. Thick (thin) lines correspond to a variation of the rate by a factor of five up (down), respectively. Lower panel: impact of the \nap~reaction rate from present work when
     the upper and lower limits of the rate are used (thick and thin lines, respectively). In both panels the uncertainty range for $^{13}$C abundances is highlighted in light blue.}
\end{figure}

The post-SN abundances have been calculated using the PPN NuGrid Post-Processing Nucleosynthesis code \cite{pignatari:16}
with the following nuclear network setup. We used 5195~species (from H to Bi, including all the unstable isotopes by $\beta$-decay with a half life longer than 10$^{-5}$~s) and 66953~reactions. We refer to \cite{pignatari:16} for a detailed list of all nuclear rates used in the network. 
For each trajectory, we ran three sets of simulations using the \nap~reaction rate from CF88 compilation, and the CF88 rate divided and multiplied by a factor of five.
The isotopic abundances profiles for the stable isotopes H, $^{4}$He, $^{12}$C, $^{13}$C, $^{14}$N, $^{15}$N and $^{16}$O, including the decay of
unstable species, and for the short-lived isotopes $^{22}$Na and $^{26}$Al are shown in Fig.~\ref{f:massFrac}, upper panel.
These are the same calculation as performed in Fig.~\ref{f:impact}, but using a set of explosive He-burning trajectories that covers the complete range of relevant temperature conditions, as described above.  
Therefore the results obtained in Fig.~\ref{f:massFrac} are consistent with Fig.~\ref{f:impact}, since the stellar conditions in the two calculations are the same. 
The only apparent difference is that while simulations based on mass coordinate refer to the specific progenitor model used, the calculations shown in Fig.~\ref{f:impact} are representative of explosive He-burning conditions independently of the original model. Therefore, the abundance profiles with respect to the SN peak temperatures is comparable to nucleosynthesis results shown with respect to mass coordinate from any model of CCSN explosive He-burning layers.
We then performed a second set of calculations, using
the low and the high thermonuclear reaction rates, from the present work, given in
Tab.~\ref{t:rates}. For comparison, the abundances obtained using these rates are shown
in Fig.~\ref{f:massFrac}, lower panel. 

In both cases, the largest impact of the \nap~reaction rate on $^{13}$C abundances is for a peak temperature of 0.54~GK. As expected the largest abundance variation decreases when the rates from the present work are used. Furthermore, the temperature range where the \nap~reaction rate has an impact is also reduced. With the \nap\ reaction rates from the present work the uncertainty on the integrated $^{13}$C yield, highlighted in light blue in Fig.~\ref{f:massFrac} (bottom), is a factor of 7 for the lower and the upper limit compared to the adopted rate. This will improve future theoretical predictions of $^{13}$C production in CCSN models with H ingestion.

\begin{figure}[!htpb]
  \includegraphics[width=0.48\textwidth]{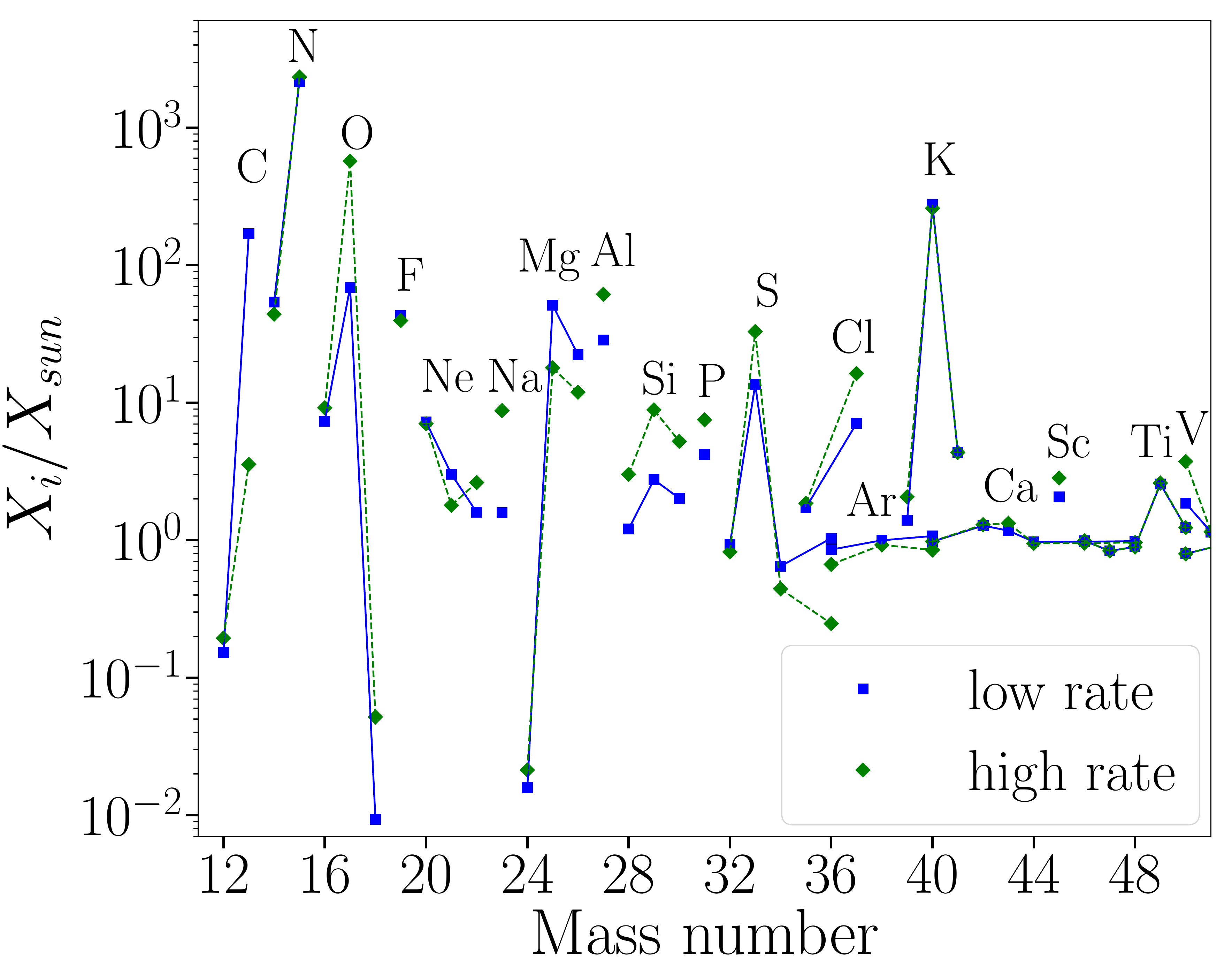}
  \caption{\label{f:prodFactor}
    (Color online) Production factors of stable isotopes, including the decay of unstable species, in the mass region between $^{12}$C and $^{50}$V, obtained using the lower limit of the \nap~reaction rate from present work (blue squares) and the upper limit (green diamonds) from the trajectory with temperature peak of the SN shock of 0.54~GK. Isotopes of a given element are connected with lines.}
\end{figure}

Fig.~\ref{f:prodFactor} also illustrates the largest impact of the \nap~reaction rate from the present work on production factors of stable isotopes, including the decay of unstable species, in the mass region between $^{12}$C and $^{50}$V, using the trajectory with the temperature peak of 0.54~GK.
From Fig. \ref{f:prodFactor} it is interesting to notice the strong impact of the \nap\ rate in making $^{13}$C and $^{17}$O during the SN shock, where the reaction is reducing the radiogenic production of $^{13}$C from the $^{13}$N decay, and favours the nucleosynthesis flow passing via $^{17}$O.
If we consider $^{17}$O for instance, a higher \nap\ rate would increase the abundance of $^{16}$O, which increases the $^{16}$O(p,$\gamma$)$^{17}$F rate, feeding the radiogenic production of $^{17}$O. In the same way, a higher \nap\ rate also increases the amount of protons available to be captured, which also increases the proton capture rate on $^{16}$O. 
Together with $^{13}$C and $^{17}$O, we find that other species affected in the He shell are between $^{23}$Na and $^{37}$Cl. This is due again to the impact that the \nap~reaction has on the $\alpha$-particle and proton budget during the SN explosion. Isotopes of the intermediate-mass elements are also produced in deeper layers of the SN ejecta, and their enhanced production in the He shell cannot be disentangled, to allow comparison with observations.

Novae and fast-rotating massive stars have been proposed as important stellar sources for $^{13}$C, $^{15}$N and $^{17}$O \citep[e.g.,][and references therein]{romano:19,chiappini:08}, but a clear picture is not yet defined. \cite{pignatari:15} discussed the possible impact in contributing to the galactic chemical evolution of $^{15}$N.  The H-ingestion in He shell layers and following nucleosynthesis in the SN shock may therefore have a strong impact on the overall production of these H-burning products. For more robust predictions for the final abundance of $^{13}$C, $^{15}$N and $^{17}$O in the type of models discussed in this work, the support of multi-dimensional hydrodynamics models is required \citep[see discussion in][]{pignatari:15}.

\section{SUMMARY AND CONCLUSIONS}
A new \nap\ reaction rate with meaningful statistical uncertainty has been 
evaluated using the most up to date \fl\ spectroscopic information. First, the 
FR-DWBA analysis of the \clt\ transfer reaction populating \ox\ states 
(analog of \fl\ states) in the $E_x = 5.6 - 7.8$~MeV range has been reported. 
Alpha spectroscopic factors were extracted and the deduced alpha-widths were 
found to be within a factor of two of reported values in the literature when 
available. Alpha spectroscopic factors were then used to deduce alpha-widths 
of \fl\ analog states when the mirror connection with \ox\ levels could be 
established. If not, assumption on the dimensionless $\alpha$-particle 
reduced widths was used ($\langle\theta^2_\alpha\rangle = 0.04$).

A Monte Carlo procedure consistently taking into account uncertainties on the 
energy, partial/total width and spin and parity of the \fl\ states was then 
used to determine the \nap\ reaction rate and its corresponding statistical 
uncertainty. Correlation effects for the energy uncertainty of \fl\ states
has been taken into account in the present work when needed.
The \nap\ nominal rate is consistent within a factor of two with previous 
rate~\cite{CF88} used in stellar models, and its uncertainty in the temperature 
range of interest is $\approx2$.
It has been shown that the main uncertainty in the reaction rate comes from 
the uncertainty associated to the alpha-width of \fl\ states. In order 
to improve this situation an experimental determination of the alpha-widths of 
unbound \fl\ states should be a priority.

The new \nap\ reaction rate and corresponding uncertainty has been used 
to study the nucleosynthesis in sixteen explosive He-burning trajectories, with temperature peaks ranging between 0.4 GK and 0.7 GK, from state-of-the-art CCSN stellar models. The abundance signature of proton ingestion in the He layer of the massive stars progenitor is considered. Results show that with the present rates the uncertainty on the $^{13}$C integrated yield from these models is about a factor of 50 when using the lower and upper reaction rates. Future stellar yields of CNO isotopes from CCSNe models including H ingestion will definitely need to consider the \nap\ reaction.

\begin{acknowledgments}
The continued support of the staff of the Tandem-Alto facility as well as the 
target laboratory staff is gratefully acknowledged. 
We thank P. Descouvemont and N. Keeley for extremely valuable discussions concerning shell-model configurations and the link with the cluster model used in DWBA.
We thank Chris Fryer and Samuel Jones for providing the trajectories used for the astrophysical simulations. 
AML acknowledges the support of the Science and Technology Facilities Council (STFC Consolidated Grant ST/P003885/1).
MP and TL acknowledges significant support to NuGrid from NSF grant PHY-1430152 (JINA Center for the Evolution of the Elements) and STFC (through the University of Hull's Consolidated Grant ST/R000840/1), and access to {\sc viper}, the University of Hull High Performance Computing Facility.
MP acknowledges the support from the "Lendület-2014" Programme of the Hungarian Academy of Sciences (Hungary).
MP and TL also acknowledge support from the ERC Consolidator Grant (Hungary) funding scheme (project RADIOSTAR, G.A. n. 724560). AML, TL and MP also thank the UK network BRIDGCE. The authors thank the ChETEC COST Action (CA16117), supported by COST (European Cooperation in Science and Technology).
\end{acknowledgments}

\bibliographystyle{apsrev4-1}
\bibliography{n13ap}

\end{document}